\documentclass[prd, 10pt,aps,nofootinbib, floatfix,notitlepage,twocolumn]{revtex4-1}
\usepackage{amsmath,amsfonts,amssymb}
\usepackage{multirow}
\usepackage{mathrsfs}
\usepackage{siunitx}
\usepackage{mathtools}
\usepackage{float} 
\usepackage{gensymb}
\usepackage[caption=false]{subfig}
\usepackage{graphicx}
\usepackage[english]{babel} 
\usepackage{hyperref} 
\usepackage{color}
\usepackage{xcolor}
\usepackage[normalem]{ulem}

\usepackage{adjustbox}

\newcommand{\be}{\begin{equation}} 
\newcommand{\ee}{\end{equation}}
\newcommand{\bes}{\begin{equation*}}
\newcommand{\ees}{\end{equation*}}

  \usepackage[utf8]{inputenc} 
\makeatletter
\let\ftype@table\ftype@figure
\makeatother

\def\mnras{\ref@jnl{MNRAS}}             
\newcommand\aap{\ref@jnl{A\&A}}
\definecolor{blanchedalmond}{rgb}{1.0, 0.92, 0.8}
\definecolor{bazaar}{rgb}{0.6, 0.47, 0.48}
\definecolor{beaver}{rgb}{0.62, 0.51, 0.44}

\begin{document}

\title{A Cosmic microwave background search for fine-structure constant evolution}
\author{
Hurum Maksora Tohfa$^{1,2,3}$}
\email{htohfa@uw.edu}
\author{Jack Crump$^{4}$}
\author{Ethan Baker$^{4}$}
\author{Luke Hart$^{5}$}
\author{Daniel Grin$^{4}$}
\email{dgrin@haverford.edu}
\author{Madeline Brosius$^{2}$}
\author{Jens Chluba$^{5}$}

\affiliation{
\small{$^{1}$Department of Astronomy, University of Washington, Seattle, Washington 98195, United States}\\
\small{$^{2}$Department of Physics, Bryn Mawr College, 101N Merion Avenue, Bryn Mawr, Pennsylvania,19010, United States}\\
\small{$^{3}$Department of Physics \& Astronomy, University of California, Riverside, Riverside, California 92521, United States}\\
\small{$^{4}$Department of Physics and Astronomy, Haverford College, 370 Lancaster Avenue, Haverford, Pennsylvania 19041, United States} and,\\
\small{$^{5}$Jodrell Bank Centre for Astrophysics, School of Physics and Astronomy, The University of Manchester, Manchester M13 9PL, UK}
}

\date{\today}

\begin{abstract}
In some extensions of the standard model of particle physics, the values of the fundamental coupling constants vary in space and time. Some observations of quasars hint at time and spatial variation of the fine structure constant $\alpha$. Here, the Bekenstein-Sandvik-Barrow-Magueijo (BSBM) model (which posits the existence of a  scalar field driving evolution in the fundamental electric charge $e$) is tested against quasar and \emph{Planck} satellite cosmic microwave background (CMB) data. In this model, variations in $e$ are coupled to the matter density through a factor {$\zeta_{\rm m}M_{\rm pl}^{2}/{\overline{\omega}}$, where $\zeta_{m}$ is the fractional electromagnetic contribution to nucleon rest masses, $\overline{\omega}$ is the energy-scale squared of new (BSBM) physics, and $M_{\rm pl}$ is the Planck energy scale}. Simulations conducted here do not support claims that the electrostatic contribution to $\zeta_{m}$ is completely shielded. Other common approximations used in BSBM {modeling} are found to be adequate. Principal components of the CMB data with respect to variations in $\alpha$ are used to obtain constraints of $\zeta_{\rm m}M_{\rm pl}^{2}/{\overline{\omega}}\lesssim 9.3 \times 10^{-9}$ for a massless field. A forecast anticipating the promise of the Simons Observatory CMB experiment shows that SO will be sensitive to values of {$\zeta_{\rm m}M_{\rm pl}^{2}/{\omega}\geq 2.2 \times 10^{-9}$, significantly improving on the uncertainty in $e$-variation from quasar spectra alone}.
\end{abstract}
\date{\today}
\maketitle
\section{Introduction}
Thanks to measurements of cosmic microwave background (CMB) anisotropies by the \emph{Planck} satellite and ground-based efforts like the South-Pole Telescope (SPT) and Atacama Cosmology Telescope (ACT), cosmic acceleration as probed by type Ia supernovae, and clustering/lensing of galaxies, there is a standard cosmological model. This concordance model has relic-density parameters of $\Omega_b h^
2 = 0.0224\pm 0.0001$, $\Omega_c h^{2}= 0.1200\pm0.0012$,
and $\Omega_{\rm DE}$ = $0.68.547\pm0.0073$ (where $b$ denotes baryons, $c$ denotes cold dark matter, or CDM, and DE represents dark energy) \cite{Benson:2014qhw,Frieman:2008sn,SupernovaCosmologyProject:2008ojh,Planck:2018vyg}. 

Using \emph{Planck} data and future results from nearly cosmic-variance-limited (CVL) CMB polarization experiments (e.g. CMB-S4 and the Simons Observatory or SO), the CMB can also be used to test models of dark-sector contents and interactions. Existing data could even provide evidence for an early dark-energy component \cite{Hill:2018lfx,Poulin:2018dzj,Poulin:2018cxd,Smith:2019ihp,Smith:2020rxx,Smith:2022hwi}, which could reconcile tension between late-time supernovae and CMB inferences of the Hubble constant \cite{Riess:2016jrr,Renzi:2017cbg}. The evolution of cosmological perturbations probed by the CMB is influenced by the photon diffusion length (see Refs.~\cite{Silk:1967kq,Hu:1995em} and citations therein), the redshift of the last scattering surface, and the detailed physics of recombination \cite{2010PhRvD..81h3005G,Chluba:2010ca,2011PhRvD..83d3513A, Chluba2006}. The CMB is thus also sensitive to the possibility that the fundamental ``constants" are in fact dynamical rather than constant.

The possibility that fundamental parameters like $e$, $m_{e}$, $m_{p}$, $G$, $\hbar$, and even the speed-of-light $c$  vary in time (or space) was raised by Dirac and others, who noticed that certain combinations of these parameters with units of time were numerically comparable to the age of the Universe \cite{1937Natur.139..323D,Teller:1948zz,Gamow:1967zza,Gamow:1967zz}. They posited that these ratios were equal to the age of the Universe at \emph{all} times, implying {a specific time evolution} for the fundamental parameters. Although the simplest realizations of this idea are readily ruled out on anthropic grounds, {more general} scalar-tensor theories of gravitation \cite{Brans:1961sx} and electromagnetism \cite{Bekenstein:1982eu} were then developed, as well as some UV completions of the standard model that predicted evolution of fundamental constants \cite{Isham:1970aw,Terazawa:1976eq,Terazawa:1981ga,Terazawa:2012fa}

The simplest extension of Maxwell electromagnetism that supports variation in $e$ [and thus the fine-structure constant $\alpha=e^{2}/(\hbar c)]$ while recovering the predictions of standard electromagnetism was put forward by Bekenstein \cite{Bekenstein:1982eu}. This theory relies on a new scalar $\psi$, which modulates the Maxwell Lagrangian for electrodynamics and has a Brans-Dicke kinetic term with coupling $\omega$. This model was extended to include the gravitational interactions of $\psi$ in Refs. \cite{Sandvik:2001rv,Barrow:2002zh}. Subsequent work (see Ref. \cite{2017RPPh...80l6902M} for a review) generalized the scalar field to a massive one with non-trivial field dependence in $\omega$ \cite{Barrow:2008ju,Barrow:2011kr}, {included} spatial perturbations \cite{Barrow:2002db,Mota:2004mz}, and developed more extensive modeling of the theory's dynamics \cite{Barrow:2002zh,Mota:2003tm,Barrow:2005sv,Barrow:2013uza}.

Broadly, varying fundamental parameters occur in standard-model extensions with extra dimensions (beginning with Kaluza-Klein theory
\cite{1926ZPhy...37..895K} and continuing with string-inspired ideas like the runaway dilaton \cite{Chodos:1979vk,Marciano:1983wy,Kolb:1985sj,Damour:1994zq,Damour:2002mi,Damour:2002nv,Overduin:1997sri,Brax:2002pf,Palma:2003rs}). Other cases include disformal theories, in which radiation and matter geodesics are given by different metrics
\cite{vandeBruck:2015ida,vandeBruck:2015rma}. {These scalars could} constitute dark energy \cite{Wetterich:1987fm,Copeland:2003cv,Parkinson:2003kf,Graham:2014hva} or dark matter (DM) \cite{Olive:2001vz,Olive:2007aj,Stadnik:2014tta,Stadnik:2015kia} with novel astronomical and lab phenomenology resulting, such as evolution in the cosmic equation of state, laser interferometry signals, and violations of the weak equivalence principle
\cite{Uzan:2002vq,2017RPPh...80l6902M,Martins:2017qxd}.

There have been hints of variation in $\alpha$ from high-resolution spectra of redshift $z\sim 3$--$7$  quasistellar objects (QSOs) \cite{Webb:1998cq,Murphy:2000nr,Chiba:2006xxm,Webb:2010hc,King:2010dwp,King:2012id}, with disputes about the interpretation of these results \cite{Molaro:2007kp,Molaro:2013saa,Songaila:2014fza,Kotus:2016xxb,Murphy:2016yqp,Murphy:2017xaz}. Current/future {observations} promise unprecedented sensitivity to $\alpha$ variations with the potential to resolve these disputes
\cite{Schmidt:2020ywz,Wilczynska:2020rxx} (e.g., {results from the ESPRESSO spectrograph, which impose} the constraint $\Delta \alpha/\alpha\sim 10^{-6}$\cite{Murphy:2021xhb}).

Any theory of varying $\alpha$ will also make a prediction at $z\simeq 1100$, the epoch of CMB decoupling. The Thomson scattering rate scales as $\alpha^{2}$, while the  hydrogen $2s\to 1s$ transition rate (the bottleneck for recombination) scales as $\propto \alpha^{8}$. The redshift and width of the last-scattering surface are influenced by model parameters \cite{Silk:1967kq,Hu:1994uz,Hu:2001bc,Story:2012wx}. CMB temperature/polarization measurements can be used to probe variations in $\alpha$, as shown by the BOOMERanG/WMAP/\emph{Planck}/SPT/ACT upper limits of Refs \cite{Battye:2000ds,Avelino:2001nr,Martins:2002iv,Rocha:2003fw,Martins:2003pe,Ichikawa:2006nm,Menegoni:2009rg,Menegoni:2009rg,Menegoni:2012tq,Martins:2010gu,Menegoni:2012tq} and forecasts of Refs. \cite{Kaplinghat:1998ry,Hannestad:1998xp,Rocha:2003gc}. Correlations with other fundamental constants (e.g., $G$) were considered in Refs. \cite{Martins:2003pe,Martins:2010gu,Menegoni:2012tq,Ichikawa:2006nm,Menegoni:2009rg}. All these analyses assumed a single nonstandard $\alpha$ value at early times. Evolution of $G$ was constrained in Refs. \cite{Hojjati:2011xd,Hojjati:2013xqa,Koyama:2015vza,Pogosian:2016pwr,Espejo:2018hxa}. 

Simply parameterized time evolution in $\alpha$ and  connections to the Hubble tension were explored in Refs.~\cite{2018MNRAS.474.1850H,Hart:2019dxi}. Given the range of theoretical possibilities and observational controversy, we {use principal component analysis (PCA), a model-independent technique in which the eigenvectors of the information matrix are found}. PCA may be used to test data for novel physics, even without a compelling model, and has been applied to explore dark matter annihilation, nonstandard recombination, and late-time cosmic acceleration \cite{Wasserman:2001ng,Huterer:2002hy,Hu:2003gh,Mortonson:2007hq,Samsing:2012qx,2012PhRvD..85d3522F,Miranda:2017mnw,DiazRivero:2019ukx}.

Models with large projections onto these eigenvectors will be best constrained by the data. Any model can be constrained through projection onto principal components (PCs), without the need to rerun a full Monte Carlo Markov chain (MCMC). PCs can elucidate the epochs driving the constraint. {Techniques for probing cosmic recombination with PCA were developed and applied in Refs.~\cite{Farhang:2011pt,2020MNRAS.495.4210H}}. In Ref.~\cite{Hart:2021kad},  PCA was applied to to obtain the constraint of $\Delta \alpha/\alpha =0.0010\pm0.0024$.

Causality requires that $\alpha$ variation occur in space if it occurs in time. The resulting non-Gaussian statistics \cite{Sigurdson:2003pd} were used to impose $\sim 10^{-2}$ level  constraints to spatial $\alpha$ variation \cite{OBryan:2013nip,Smith:2018rnu}. Secondary CMB sight lines to galaxy clusters constrain variation in $\alpha$ \cite{deMartino:2016tbu} because of the {Sunyaev-Zeldovich (SZ)} effect \cite{Sunyaev:1980vz,Sunyaev:1980nv}, {although these inferences remain challenging due to relativistic corrections \citep{2022MNRAS.517.5303L}. Large volumes will be mapped by upcoming observations of neutral hydrogen at high redshift \cite{Pritchard:2011xb}, probing varying-$\alpha$ theories \cite{Khatri:2007yv,Lopez-Honorez:2020lno}.

Here, we model the evolution of $\psi$ and $\alpha$ dynamically in the Bekenstein-Sandvik-Barrow-Magueijo (BSBM) model. We allow for $\psi$ to have a mass $m$, motivated by recent work finding that light scalars and pseudoscalars are numerous in string-theory realizations and possibly cosmologically important \cite{Svrcek:2006yi,Arvanitaki:2009fg,Cicoli:2012sz,Hlozek:2014lca,Marsh:2015xka,Drlica-Wagner:2022lbd}.

We use the PCA decomposition for $\alpha$ variation and constraints from \emph{Planck} data in Ref.~\cite{Hart:2021kad} to test the BSBM model, determining the allowed values for the coupling $\zeta_m/\omega$ (where $\zeta_m$ is the ratio of nuclear electromagnetic to total rest-mass energy) and $m$. Our constraint (at $95\%$ C.L) is $\zeta_{m}/\omega\leq 9.3\times 10^{-9}$ for $m=0$, with constraints  relaxing for $mc^{2}\gtrsim 1.4 \times 10^{-32}~{\rm eV}$.

We thus obtain some of the first constraints to the BSBM model that include the full time dependence of the model, going back to the recombination era.\footnote{In the preparation of this work, we became aware of other recent work that applies CMB data to the BSBM model \cite{Martins:2022pue,Vacher:2023gnp}. These results are complementary, as we allow for the possibility that $\psi$ is massive, use PCA methods, and obtain forecasts for future CMB experiments. We also obtain QSO prior-free results in our CMB-only analysis.} The next decade of CMB measurements will bring an order of magnitude improvement in precision, and the possibility of testing many novel physical scenarios for the dark sector, neutrinos \cite{Abazajian:2016yjj,SimonsObservatory:2018koc}, and varying $\alpha$. We also apply conduct a forecast for the sensitivity of the SO \cite{SimonsObservatory:2018koc} to test the BSBM model. We obtain a $\zeta_{m}/\omega \leq 2.2 \times 10^{-9}$ sensitivity forecast. Results for $m\neq 0$ are presented in the body of the paper. 

In the BSBM model, $\psi$ couples to the trace of the radiation stress-energy tensor ($\propto$ ${E}^{2}-{B}^{2}$), where ${E}$ and ${B}$ are electric and magnetic field amplitudes. This trace is nontrivial to calculate \cite{Gasser:1982ap,Livio:1998pp,Kraiselburd:2009uh,Kraiselburd:2011ac,Kraiselburd:2018uac,Will:2018bme,Bekenstein:2002wz}. Observational probes of BSBM typically assume ${E}^{2}-{B}^{2} \propto \rho_{\rm m}$, the total cosmological matter density, with a proportionality constant $\zeta_{m}$. Estimates of $\zeta_{m}$ vary in sign and in magnitude by $2-3$ decades \cite{Gasser:1982ap,Livio:1998pp,Kraiselburd:2009uh,Kraiselburd:2011ac,Kraiselburd:2018uac,Will:2018bme,Bekenstein:2002wz}, depending on whether it is dominated by electric or magnetic contributions \cite{Bekenstein:2002wz}, classical vs quantum approaches, and our ignorance of the dark-sector contribution \cite{Gasser:1982ap,Livio:1998pp,Kraiselburd:2009uh,Kraiselburd:2011ac,Kraiselburd:2018uac,Will:2018bme,Bekenstein:2002wz}.

We ran a simulation to test recent claims \cite{Bekenstein:2002wz} that magnetic contributions to $\zeta_{m}$ dominate over electrostatic terms. Our results indicate that they do not. We then put constraints on $\zeta_{m}/\omega$. Most comparisons of the BSBM model to data rely on a non-energy-conserving approximation in the equations of motion. We find that our constraints are insensitive to this approximation.

We begin with a discussion of scalar field dynamics and numerical methods for their evolution in Sec.~\ref{sec:bsbm_theory}. We review principal component methods and explain our use of them in Sec. \ref{sec:pca}. We summarize our MCMC modeling techniques and results in Sec. \ref{sec:results}, and conclude in Sec.~\ref{sec:conclusions}. In Appendix~\ref{sec:zeta}, we test the cancellation of electrostatic contributions to $\zeta_m$. In Appendix~\ref{sec:bsbm_subtle}, we explore the impact of an energy nonconserving approximation made throughout the literature to solve the BSBM equations of motion, and find it to be negligible.  

\section{BSBM theory}
\label{sec:bsbm_theory}
In Bekenstein's original formulation of this $\alpha$ variation theory \cite{Bekenstein:1982eu}, the scalar field $\psi$ was coupled to the standard Maxwell Lagrangian, but the coupling to gravity was not included. Later work by Sandvik, Barrow, and Magueijo included this inevitable coupling \cite{Sandvik:2001rv}. In the resulting Bekenstein, Sandvik, Barrow, Magueijo (BSBM) theory \cite{Sandvik:2001rv,Bekenstein:2002wz}, the electric charge $e$ evolves, while Planck’s constant $h$ and the speed of light $c$ are constant. A real scalar field modulates the Maxwell Lagrangian, causing variations in $\alpha$:  $e_{0} \rightarrow e=e_{0} \epsilon\left(x^{\mu}\right)$. Here  $\epsilon$ is a dimensionless scalar field. Then $\epsilon$ couples to the electromagnetic gauge field $A_{\mu}$ in the Lagrangian. Under the usual local gauge transformation $U(x) = e^{i \theta(x)}$, the electromagnetic action is still invariant with a modified gauge-field transformation, $\epsilon A_{\mu}\to \epsilon A_{\mu}+\partial_{\mu}\theta(x)$.
 
Recasting the equations in terms of a more standard scalar field $\psi$ defined by $\psi\equiv \ln{(\epsilon)}$, the equation of motion for the scalar field is \cite{Bekenstein:2002wz}:
\begin{equation}
\label{eq:generic_bsbm}
\square \psi=\frac{2}{\overline{\omega}} e^{-2 \psi} \mathcal{L}_{\mathrm{e m}},
\end{equation}
where $\overline{\omega} = \hbar c/ l^2$ is a coupling constant and $l$ is some new length scale below which Coulomb's law breaks down. In Planck units, $\overline{\omega}$ has units $[{\rm energy}]^2$, and so we employ the reparametrization $\overline{\omega}=M_{\rm pl}^{2}\omega$ in terms of the reduced Planck mass $M_{\rm pl}=1/\sqrt{8\pi G}$ and a dimensionless parameter $\omega$ that quantifies the amplitude of evolution in $\alpha$.  

The standard Maxwell Lagrangian, $\mathcal{L}_{\text{em}}=F_{\mu \nu}F^{\mu \nu}/4$, vanishes for pure radiation (as $\mathcal{L}_{\rm em}\propto E^{2}-B^{2}=0$, where $E$ and $B$ are the electric and magnetic-field vector amplitudes, respectively). It would, however,  be excited by plasma-sourced electrostatic and magnetic fields $\mathbf{E}$ and $\mathbf{B}$ in the early Universe. The approximation $\mathcal{L}_{\rm} \simeq \zeta_{m}\rho_{m}$ is used throughout the literature, where $\zeta_{m}$ is a dimensionless constant quantifying the electromagnetic self-energy of nonrelativistic matter. This replacement warrants justification, and requires a numerical estimate of $\zeta_{m}$.

\subsection{Values of $\zeta_m$}
We summarize past results and quantitatively assess  claims \cite{Bekenstein:2002wz} that prior calculations of $\zeta_{m}$ were in severe error, using a numerical plasma simulation in Appendix \ref{sec:zeta}. The simplest possibility is that the dark sector does not source scalar-field evolution. Straightforward estimates of the baryonic contribution to $\zeta_{m}$ may be obtained by extrapolating a semiempirical mass formula for nuclear electromagnetic self-energy to the case of a primordial (hydrogen + helium) plasma \cite{1977_segre,1998_Lang}:
 \begin{equation}
     E_{\text{Coulomb}}=-a_C\frac{Z^2}{A^{1/3}}.
 \end{equation}
 
In this equation, $a_C$ is an empirically determined constant, $Z$ is the atomic number of the nucleus, and $A$ is the mass number of the nucleus. This term represents the decrease in nuclear binding energy that is caused by the electrostatic repulsion between the positively charged protons in the nucleus. 

Using the fact that $a_C\approx 0.7$ MeV \cite{Bekenstein:1982eu}, the estimate 
$\zeta_{m}\sim 1.3\times10^{-2}$ was obtained through classical approximations to the nucleus. Late-time nucleosynthesis could lead to additional time dependence \cite{Bekenstein:1982eu,Livio:1998pp}. These estimates are dominated by $E^{2}$ (rather than $B^2$). Subsequent estimates modeled quark electromagnetic fields within the nucleus \cite{Sandvik:2001rv}, applying the Born term in the Cottingham formula and following the methods of 
Ref.~\cite{Gasser:1982ap} to obtain $\zeta_m \approx 10^{-4}$. 

As $\rho_{m}$ is dominated by dark matter (whose nature is unknown), $\zeta_{m}$ could be dominated by nonbaryonic contributions \cite{Sandvik:2001rv}. One can then have $-1<\zeta_{m}<1$ [with the bounds saturated for dark matter composed of superconducting strings]. A negative value or $\zeta_{m}$ is appealing because it can explain QSO results hinting at a decrease of $\alpha$ with time.

Subsequently \cite{Bekenstein:2002wz}, the author of Ref.~\cite{Bekenstein:1982eu} solved the modified Maxwell equations of BSBM theory and the static limit of Eq.~(\ref{eq:generic_bsbm}), claiming that (under some approximations) the $\psi$ field configuration shields out the electrostatic contribution  to Eq.~(\ref{eq:generic_bsbm}), leaving only the $B^{2}$ term.

This cancellation is also claimed to allow the BSBM theory to evade terrestrial, Solar system constraints to weak-equivalence principle (WEP) violation \cite{Bekenstein:2002wz}, although the magnetic term might also lead to detectable WEP violation \cite{Kraiselburd:2009uh}. More generally, WEP bounds may be satisfied for $|\zeta_{m}|>10^{-3}$ for the $\zeta_{m}/\omega$ values that saturate the CMB constraints of Sec.~\ref{sec:results}, motivating us to continue testing this theory empirically.

If this cancellation occurs, then the magnetic contributions from baryons dominate, and one can have negative $\zeta_{m}$ without exotic dark sector contributions of the type discussed in Refs.~\cite{Sandvik:2001rv,Olive:2001vz}. Reference~\cite{Bekenstein:2002wz} provides estimates of $\zeta_{m}$, integrating the classical solution for ${B}^2/8\pi$ outside of the Compton radius of a proton, weighting quantities by the abundances of hydrogen and helium in the Universe, and obtaining the estimate $\zeta_{m}\approx -1.98\times 10^{-5}$. This estimate does not apply the effective field theory methods of Ref.~\cite{Gasser:1982ap}.

Accounting more accurately for the composite and quantum mechanical nature of the nucleus, Ref.~\cite{Kraiselburd:2009uh} applies methods introduced in Refs.~\cite{1977PhRvD..15.2711H,Will:2018bme} to compute  $\zeta_{m}$ in terms of expectation values of nuclear electromagnetic density and current operators. Quantum mechanical identities (e.g. the Thomas-Reich-Kuhn rule) can be used to rewrite these quantities in terms of the photoabsorption cross section and other empirically measurable properties of nuclei. An estimate of 
 \begin{equation}
    \zeta_{m}(A)\approx -\frac{8.60465\times10^{-6}}{A^{1/3}}
\end{equation} is obtained \cite{Kraiselburd:2011ac}, although this calculation does not yet apply the fully relativistic field-theory techniques of Ref.~\cite{Gasser:1982ap}.

Seeking to test the analytic approximations of Ref.~\cite{Bekenstein:2002wz}, we conducted a simulation of the early Universe plasma to assess if the electrostatic contribution to $\zeta_{m}$ is shielded. Our methods and results are presented in Appendix \ref{sec:zeta}. We find that the term of opposite sign potentially driving cancellation in $\zeta_{m}$ is orders of magnitude smaller than other terms, and thus that electrostatic cancellation does not occur. Further investigation is needed, but we recommend the use of $\zeta_{m}\approx 10^{-4}$ as a ``standard" value for the BSBM variant in which {$\psi$ is not coupled to} the dark sector.

Broadly, there are two logical possibilities. In one case, {$\alpha$ variation is sourced by a coupling of $\psi$ to baryonic matter.} In the other case, {$\alpha$ variation is sourced by a coupling of $\psi$ to the dark sector.}
In the case that  $\zeta_{m}$ is sourced only by standard electromagnetism, {it is important to compute $\zeta_m$ accurately}, {so that measurements of or limits to $\alpha$ variation can be turned into allowed ranges for $\sqrt{\overline{\omega}}$, the energy scale of new physics. Put another way, an accurate value of $\zeta_m$ is essential to go from the limits obtained in Sec.~\ref{sec:results} to an empirically allowed range for $l=\hbar c/\sqrt{\overline{\omega}}$, the length scale at which Coulomb's law breaks down due to interactions of standard-model fields with the BSBM scalar field.} 

In the {case that $\alpha$ variation is sourced by dark-sector interactions of $\psi$}, $\zeta_{m}M_{\rm pl}^{2}/\overline{\omega}$ is a single effective (unknown) dimensionless coupling constant (e.g. $g$ in a dilaton warp factor of the form $e^{-g \psi}$ in a nonminimal coupling term of a Lagrangian) to be determined empirically or predicted in an effective dark sector theory, as discussed extensively in  Ref.~\cite{Olive:2001vz}. There, values of $\zeta_{m}M_{\rm pl}^{2}/\overline{\omega}=-\sqrt{2}/4$, $1/2$, $0.05$, and $-1/\sqrt{16 \Omega+24}$ are obtained, for the string dilaton, supersymmetric Bekenstein-Magueijo model, gaugino-driven modulus, and Brans-Dicke electromagnetism model variants, respectively. Here $\Omega$ is the usual Brans-Dicke coupling parameter. For the remainder of this paper, we treat $\zeta_{m}M_{\rm pl}^2/\overline{\omega}$ as a parameter to be empirically determined from cosmological data.

\subsection{Scalar field dynamics}
\label{sec:sf_dynamics}
The evolution of the scalar field from Eq.~(\ref{eq:generic_bsbm}) can {then} be written
\begin{equation}
\label{eq:time_dependent}
\ddot{\psi}+3 H \dot{\psi}=-\frac{2}{\overline{\omega}} e^{-2 \psi} \zeta_{m} \rho_m .
\end{equation}

\begin{figure*}[ht]
\includegraphics[width=\textwidth]{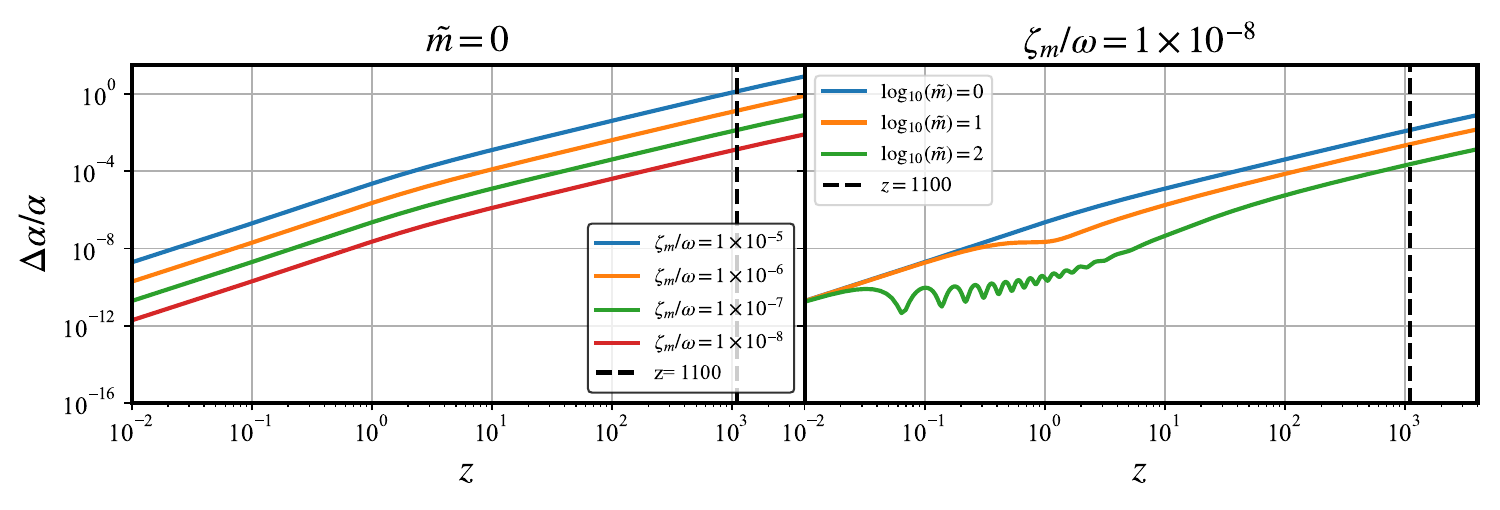}
\caption{Evolution of $\alpha$ variation as a function of $z$ for a range of BSBM model parameters. }
\label{fig:alpha}
\end{figure*}

Assuming a spatially flat, homogeneous, and isotropic Universe, we write Eq.~(\ref{eq:time_dependent}) as a function of redshift $z$ using the standard Hubble parameter definition $H\equiv \dot{a}/a$ where $H=H_0 \sqrt{\Omega_m (1+z)^3+\Omega_r (1+z)^4+\Omega_{\Lambda}}$ and standard relation between scale factor $a$ and $z$, $a =1/(1+z)$:
\begin{widetext}
\begin{align}
\label{eq:bsbm_gov_eq}
\frac{d^{2} \psi}{d z^{2}} -& \frac{d \psi}{d z}\left\{\frac{2}{(1+z)}-\frac{d\ln{E(z)}}{dz}\right\} + \frac{ \tilde{m}^2  \psi}{ (1+z)^2 \left[\Omega_m (1+z)^3+\Omega_r (1+z)^4+ \Omega_{\Lambda}\right] } \nonumber \\ =& - \frac{ 6\zeta_m}{\omega} \frac{\Omega_{m} (1+z)}{\sqrt{\Omega_m (1+z)^3+\Omega_r (1+z)^4+\Omega_{\Lambda}}}e^{-2 \psi},
\end{align}
\end{widetext}where $E(z)\equiv H(z)/H_{0}$.

Here we also allow a mass $m$ for the scalar field to allow a more general set of dynamical variation models for $\alpha$. In the Klein-Gordon equation with MKS units, one has a term
$=(mc^2/\hbar)^2\psi$. After rewriting the Klein-Gordon equation through the preceding series of transformations, it is straightforward to see that $\tilde{m}=mc^{2}/(\hbar H_{0})=mc^{2}
/(2.13\times 10^{-33}~{h}~{\rm eV})$, where $h$ here is the usual dimensionless Hubble constant. For the analysis in the main body of this paper, we have assumed the standard $\rho_{m}\propto a^{-3}$ scaling. As discussed 
in Appendix \ref{sec:bsbm_subtle}, this does not self-consistently allow for energy conservation under the $\mathcal{L}_{\rm em}\to \zeta_{m}\rho_{m}$ approximation. 
We assess this issue quantitatively in Appendix \ref{sec:bsbm_subtle}, and find that the resulting correction to our field evolution, observables, and constraints is negligible.

The coefficient on the rhs of Eq.~(\ref{eq:bsbm_gov_eq}) is defined relative to the reduced Planck mass and thus appears different from that shown in some other work (e.g. Ref. \cite{Martins:2022pue}) on this topic. We treat the scalar-field coupling $\zeta_m/\omega$ and $\tilde{m}$ as free parameters to be empirically constrained. The scalar field, $\psi$ is related to $\alpha$ by $
\alpha=e^{2\psi}e_{0}^{2} /hc $, implying that $\Delta \alpha/ \alpha \simeq 2 \psi$ {for small $\psi$}.

We solved Eq.~(\ref{eq:bsbm_gov_eq}) going from $z=0$ to higher values, using an $8^{\rm th}$-order Runge-Kutta (RK) method \cite{10.2307/2156139} with $80000$ linear steps in $z$. The number of steps was increased until the relative fractional numerical error in $\Delta \alpha/\alpha$ was smaller than  $\sim 10^{-7}$, the  fractional $\alpha$ variation implied ($\sim 10^{-5}$) by some QSO measurements \cite{Sandvik:2001rv}. Note that this calculation itself does not depend on the QSO measurement. All the high-$z$ behavior is determined by Eq.~(\ref{eq:bsbm_gov_eq}), along with the chosen values of $\zeta_{m}/\omega$, $\psi(z=0)$ and $\psi'(z=0)$. Here the $'$ denotes a derivative with respect to $z$.

To verify the stability and accuracy of this method, we ran a test in which we evolved our numerical solver backward in time to high $z$ and used $\psi(z_\text{high})$ and $\psi'(z_\text{high})$ as the initial conditions for a forward integration in time (decreasing $z$), comparing $z=0$ and intermediate values of $\psi$ and $\psi'$ with the results obtained from the high-to-low $z$ evolution. We recovered the desired fractional $\sim 10^{-7}$ precision in $\Delta \alpha/\alpha$. Assuming that 
$\psi=0$ and $\psi'=0$ at $z=0$ (needed for consistency with present-day lab constraints), we show the resulting cosmological $\alpha$ variation as a function of $z$ in Fig.~\ref{fig:alpha} for a variety of $\zeta_m$ and $\tilde{m}$ values. 

Reference \cite{2017RPPh...80l6902M} summarizes Keck HIRES QSO observations  hinting at $\alpha$ variation. There, a slightly different convention for BSBM constants is used. Effectively, they constrain $\tilde{\zeta}=8\pi \zeta_{m}/\omega$, obtaining $\tilde{\zeta}\leq 3.7 \times 10^{-6}$ at $95\%$ C.L. and $\zeta_{m}/\omega\sim 10^{-7}$. In Fig.~\ref{fig:alpha}, we see that if these constraints are saturated, one expects $\Delta \alpha/\alpha\sim 10^{-2}$ at the recombination epoch, well within reach of the CMB's sensitivity to $\alpha$ variation \cite{Battye:2000ds,Avelino:2001nr,Martins:2002iv,Rocha:2003fw,Martins:2003pe,Ichikawa:2006nm,Menegoni:2009rg,Menegoni:2009rg,Menegoni:2012tq,Martins:2010gu,Menegoni:2012tq,Hart:2019dxi}. For values ${\rm \log}_{10}{\tilde{m}}\gtrsim 1.5$, the oscillations in $\psi$ are fast enough that many (tens or more) have occurred in a period comparable to the Hubble timescale today. 

In this regime, we use an adaptive version of the RK8 solver (step size is set by comparing the value for the nonadaptive regime with $\sim 1/10$ of a Hubble time, and choosing the minimum). We verified that field evolution agrees between fixed and adaptive step size solvers at the $\sim 5\%$ level for ${\tilde{m}}= 1.0$. We have verified that constraints to $\zeta_{m}/\omega$ are insensitive to an increase of time resolution (by a factor of $\sim 2$) at the $\sim 1-5\%$ level depending on the precise value of ${\tilde{m}}$.

\section{Principal Component Analysis}
One powerful approach for probing non-standard physics is PCA, which eschews a specific model, and instead determines a non-parametric family of template functions. These PCs are data-driven models that capture the variance between observations and a fiducial (or best-fit) model. 

PCA has already been used to great effect to test models of dark energy [as parametrized by its equation-of-state parameter $w(z)$] \cite{Huterer:2002hy,Miranda:2017mnw,DiazRivero:2019ukx}, the cosmic reionization history \cite{Mortonson:2007hq,Hu:2003gh}, nonstandard cosmic recombination \cite{Farhang:2011pt}, as well as more exotic physics like dark-matter decay and annihilation \cite{2012PhRvD..85d3522F}. In this work, we use the $\alpha$-variation in PCs already determined in prior work by some of us \cite{2020MNRAS.495.4210H,Hart:2019dxi} to probe time-varying $\alpha$, setting the stage for the work described here, as well as future analysis of a broad family of theoretical models. Full details of the techniques used to apply and constrain the PCs can be found in Refs.~\cite{2020MNRAS.495.4210H,Hart:2019dxi}, but we briefly review the technique below.

PCs are obtained by diagonalizing the Fisher information matrix \cite{Tegmark:1996bz,Hu:1997mj}:\begin{equation}
F_{i j}=\left\langle\frac{\partial^{2} \mathcal{L}}{\partial \theta_{i} \partial \theta_{j}}\right\rangle,
\end{equation} where $\mathcal{L}$ is the log-likelihood function of a data set or simulation given values of all model parameters $\theta_{i}$ evaluated at their fiducial values. Schematically, the parameter vector $\mathbf{\theta}=\left\{\mathbf{p},\mathbf{q}\right\}$, where $\mathbf{p}$ denotes fiducial model parameters and $\mathbf{q}$ denotes the expansion coefficients of non-standard deviations from the fiducial model for quantities (e.g. $\alpha$) usually treated as constant. 

For our case, this means that
\begin{equation}
    \frac{\Delta \alpha\left(z\right)}{\alpha_{0}}=\sum_{i}q_{i}f_{i}(z),
\end{equation}where the basis functions $f_{i}$ are some (complete, but not necessarily orthogonal) set of smooth basis functions centered at some set of redshifts $z_{i}$ and $q_{i}$ are expansion coefficients. The vector $\mathbf{p}=\left\{A_{s},n_{s},\Omega_b h^{2},\Omega_{c}h^{2}, \tau_{\rm reion},H_{0}\right\}$ contains the standard cosmological parameters of the dimensionless amplitude of the primordial perturbation power spectrum, its spectral index, the relic baryon density, CDM density, optical depth to reionization, and Hubble constant, respectively. 

For a Gaussian likelihood, the CMB Fisher matrix is given by 
\begin{equation}
    F_{ij}=\sum_{\ell}f_{\rm sky}\left(\frac{2\ell+1}{2}\right)\frac{\partial\mathbf{C}_{\ell}}{\partial \theta_{i}}\mathbf{\Sigma}_{l}^{-1}\frac{\partial\mathbf{C}_{\ell}}{\partial \theta_{j}},
\end{equation} where $\mathbf{C}_{\ell}=\left\{C_{\ell}^{\rm TT}, C_{\ell}^{\rm EE}, C_{\ell}^{\rm TE}\right\}$ is the theoretically predicted set of CMB (temperature/polarization auto, and cross) power spectra, $\mathbf{\Sigma}_{l}$ is the covariance matrix of observationally estimated CMB power spectra at multipole index $\ell$, including the effect of instrumental noise and cosmic variance, and $f_{\rm sky}$ is the fraction of sky covered by an experiment. 

Using a set of Gaussian basis functions, a modified version of the \textsc{camb}  \cite{camb} code interfaced with the \textsc{CosmoRec} recombination code \cite{2013ascl.soft04017C},
 the \emph{Planck} 2018 likelihood function, and the usual analytic approximations for the instrumental properties of the SO (an ongoing ground-based CMB experiment), a set of PCs for time-varying $\alpha$ was obtained in Ref.~\cite{Hart:2021kad}. \textsc{camb} was run including the impact of gravitational lensing \cite{2010GReGr..42.2197H}, which smooths the high-$\ell$ anisotropies. We apply these PCs here to test the BSBM model.

Several PCs are shown for the SO case in Fig.~\ref{fig:pc_show}. {These PCs peak in the range $800\lesssim z\lesssim 1200$. As described in Refs.~\cite{2018MNRAS.474.1850H,2020MNRAS.495.4210H,Hart:2019dxi,Hart:2021kad} (and references therein), evolution in $\alpha$ affects the CMB by modulating the Lyman-$\alpha$ absorption cross section, the hydrogen two-photon $2s\to 1s$ transition rage (both of which impact the free electron fraction), and the Thomson scattering cross section. All these rates in turn affect the median redshift and width of the last-scattering surface, the rate of diffusion damping, and the efficiency with which CMB polarization is generated. All these processes occur primarily during the peak $z$-range of the PCs.}

We see that {at low $\tilde{m}$}, the BSBM model is relatively featureless compared to the PCs in the $z$ range of interest. Models with {higher $\tilde{m}$ values show an overlap of large amplitude $\alpha$ variation and oscillation, increasing the importance of higher-index PCs. Perhaps more complicated potential energy functions (which likely begin coherent oscillation earlier) would show an overlap of large amplitude $\alpha$ variation and oscillation, leading to more interesting interaction with the PCs. We will explore this possibility more in further work.}

In terms of the PCs $E_{i}(z)$, any model may be expressed
\begin{equation}
    \frac{\Delta \alpha(z)}{\alpha_{0}}=\sum_{i}\rho_{{\rm M},i}E_{i}(z),
\end{equation} where
the PCs can be expressed as linear combinations of the original basis functions
\begin{equation}
    E_{i}(z)=\sum_j e_{ij} f_{j}(z),
\end{equation} where $e_{ij}$ denotes the $j^{\rm th}$ basis-component of the $i^{\rm th}$ Fisher-matrix eigenvector $\mathbf{e}_{i}$. For sufficiently dense basis sets, the PCs should themselves be numerically convergent (checked for the $\alpha$-variation case in Ref.~\cite{2020MNRAS.495.4210H}) and basis-independent (checked for variations in the cosmic recombination history in Ref.~\cite{Farhang:2011pt}). 

The projection amplitudes for any specific model realization are given by 
\begin{equation}\label{eq:proj}
\rho_{{\rm M},i}=\int \frac{\Delta \alpha(z)}{\alpha_0} \times  E_{i}(z) dz,
\end{equation} and may be used to accurately reexpress the variation around the fiducial model as long as it is small.

These can be used to construct a $\chi^{2}$ diagnostic with the best-fit (from the data) values for the model-expansion coefficients in the PC basis, $\rho_{\rm {\rm D}, i}$:
\begin{equation}
    \chi^2 = \sum_{i}\left(\rho_{\text{M},i} - \rho_{\text{D},i}\right) 
\left(\mathcal{F}\right)^{-1}_{ij}\left(\rho_{\text{M},i} - \rho_{\text{D},i}\right).\label{eq:chi_squared}
\end{equation} Here the sum is over PC indices.

One advantage of this method is that even if the likelihood with respect to model (e.g. BSBM) parameters is non-Gaussian, the likelihood with respect to PC amplitudes is still very close to Gaussian. Here the covariance is given by 
\begin{equation}
    \sigma_{\rho_{{\rm M},i}}=\sqrt{\left(\mathcal{F}^{-1}\right)_{ii}}\simeq \frac{1}{\sqrt{\lambda}_{i}},
\end{equation}where $\mathcal{F}$ is the (nearly diagonal) $\alpha$-variation Fisher matrix. The quantity $\lambda_{i}$ is the $i^{\rm th}$ eigenvalue of the $\alpha$-variation Fisher matrix. {As discussed in Ref.~\cite{Farhang:2011pt}, two types of PCs are possible for new physics. In one, standard cosmological parameters are held fixed, and $\mathcal{F}$ only includes variations in $\alpha$. In the other, the Fisher matrix includes cosmological parameter variations and $\alpha$ variation (and thus their full degeneracy structure). The former are marginalized over to obtain a Fisher-matrix for $\alpha$ variation, which was then diagonalized to obtain the PCs.}

{Applying the results of Ref.~\cite{Hart:2021kad}, we use these post-marginalization PCs, whose best-fit values and allowed ranges were originally found through a full MCMC in which cosmological parameters were simultaneously varied with PC amplitudes \cite{Hart:2021kad}.} At Fisher level, expanding around the fiducial model, parameter changes (from $\rho \to \zeta_{m}/\omega$) commute with cosmological parameter marginalization, and so we do not expect our constraints to get less stringent due to any neglected covariances. Nonetheless, in future work, we will more fully account for these covariances by using the original samples from the MCMC to construct a kernel-density likelihood in which PC amplitudes and cosmological parameters can be simultaneously varied, as in e.g. Ref.~\cite{Heinrich:2016ojb}.  

To construct our \emph{Planck} posterior probability for BSBM model coefficients, we use Eqs.~(\ref{eq:proj})-(\ref{eq:chi_squared}) and the usual $\mathcal{L}\propto e^{-\chi^{2}/2}\times \pi$. {To efficiently explore a broad parameter space, we use log-flat and flat priors $\pi$ for $\tilde{m}$ and $\zeta_{m}/\omega$, respectively, and the parameter ranges $-8 \leq \log_{10}(\tilde{m})\leq 1$ and $-10^{-2}\leq \zeta_{m}/\omega\leq 10^{-2}$. In Sec.~\ref{sec:results}, we will explain a number of tests done to make sure our constraints are robust and not driven by our choice of prior.} 

We use the best-fit PC amplitudes and errors (obtained from MCMC runs) from Ref.~\cite{Hart:2019dxi}.  We computed $\Delta\alpha(z)/\alpha$ using the method discussed in Sec. \ref{sec:bsbm_theory}. To conduct forecasts for SO, we built a mock likelihood, assuming that the fiducial model is true (e.g. that $\rho_{{\rm D},i}=0$), and used Fisher-forecast errors on the PC amplitudes. For our \emph{Planck} analysis, three PCs were used, while $10$ PCs were included for SO, to allow for its higher information content.  
\begin{figure}[h]
\includegraphics[keepaspectratio=true, scale = .33]{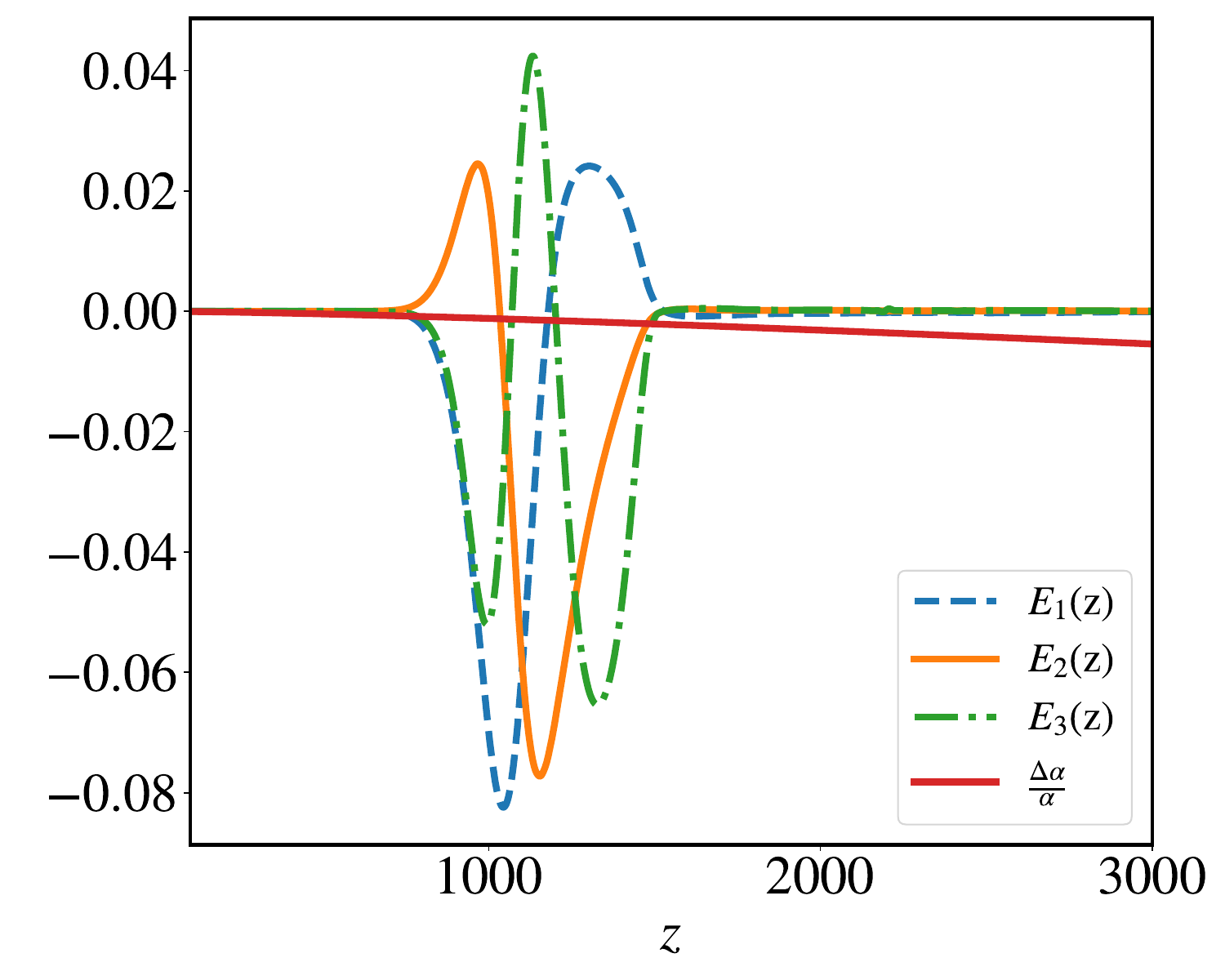}
\caption{Evolution of $\alpha$ variation as a function of $z$ assuming $\tilde{m}=0$, $\zeta_m/ \omega= 1 \times 10^{-8}$ and first three SO eigenmodes. }
\label{fig:pc_show}
\end{figure}
It can be helpful to assess the relative information content and utility of different PCs, and to determine how many are truly needed to properly capture a data set. One useful tool is the 
signal-to-noise (SNR) contribution from each mode, which is 
\begin{equation}
(S / N)_i=\sqrt{\lambda_i\rho_i ^2}.\label{eq:permode_snr}
\end{equation}

Another interesting quantity in PCA is the risk factor \cite{Wasserman:2001ng,Huterer:2002hy,Samsing:2012qx}, defined via
\begin{eqnarray}
\sigma^2\left[\alpha\left(z_j\right)\right] &=& \left(\sum_i^{N_{\rm PC}} \frac{e_i^2\left(z_j\right)}{\lambda_i}\right),
\\
b\left(z_j\right) &=& \frac{\Delta \alpha\left(z_j\right)}{\alpha}-\sum_i^{N_{\rm PC}}\left[\rho_i E_i\left(z_j\right)\right], 
\\
\operatorname{Risk}[N_{\rm PC}] &=& \sum_j\left\{b^2\left(z_j\right) + \sigma^2\left[\alpha\left(z_j\right)\right]\right\},\label{eq:risk}
\end{eqnarray} where $N_{\rm PC}$ is the number of PCs used to test a model. Here $b(z_j)$ is defined to be the bias in $\Delta \alpha(z)/\alpha$ induced by using an incomplete set of PCs, which competes with the variance (which decreases when PCs are filtered out to reduce the error in the data if the fiducial model is actually true). This quantity is model dependent and minimized when an optimal number of PCs is chosen to test a specific model. In Sec.~\ref{sec:results}, we compute this quantity to assess the impact of different PCs on our BSBM model constraints.

\label{sec:pca}

\section{Data Analysis and Results}
\label{sec:results}
Using the \emph{Planck} 2018 $\alpha$-variation likelihood described in Sec.~\ref{sec:pca} and our scalar-field integrator, we found best fit values of $\zeta_{m}/\omega$ for the $m=0$ case. We then set a broad parameter range of $-0.01\leq \zeta_{m}/\omega\leq 0.01$ and $-8\leq \log_{10}{(\tilde{m})}\leq 1$ and ran an MCMC simulation to determine the allowed $\zeta_{m}/\omega$ and $\log_{10}(\tilde{m})$ parameter space. 

We used the \textsc{emcee} package, which applies Goodman \& Weare's affine invariant sampler to run MCMCs and checked for convergence using the autocorrelation time \cite{Foreman-Mackey:2012any}.  For the CMB analysis, we established two-dimensional MCMC convergence as follows: For the \emph{Planck} analysis, we ran $32$ chains for $12000$ samples. We obtained a correlation length of $\sim 150$ for $\zeta_{m}/\omega$ and $\sim 60$ samples for $\log_{10}(\tilde{m})$, less than $12000/50=240$. For the SO analysis, we ran $32$ chains for $10000$ samples. We obtained an autocorrelation length of $\sim 10 $ samples $\zeta_{m}/\omega$ and $58$ in $\log_{10}(\tilde{m})$. Both are less than $10000/50=200$, and so all these CMB chains are converged. For our analysis, we used a burn-in fraction of $0.3$ throughout. Visualizations and confidence intervals were generated with the \textsc{GetDist} package \cite{Lewis:2019xzd}. 
\begin{figure*}[ht]
\includegraphics[keepaspectratio=true,scale=.7]{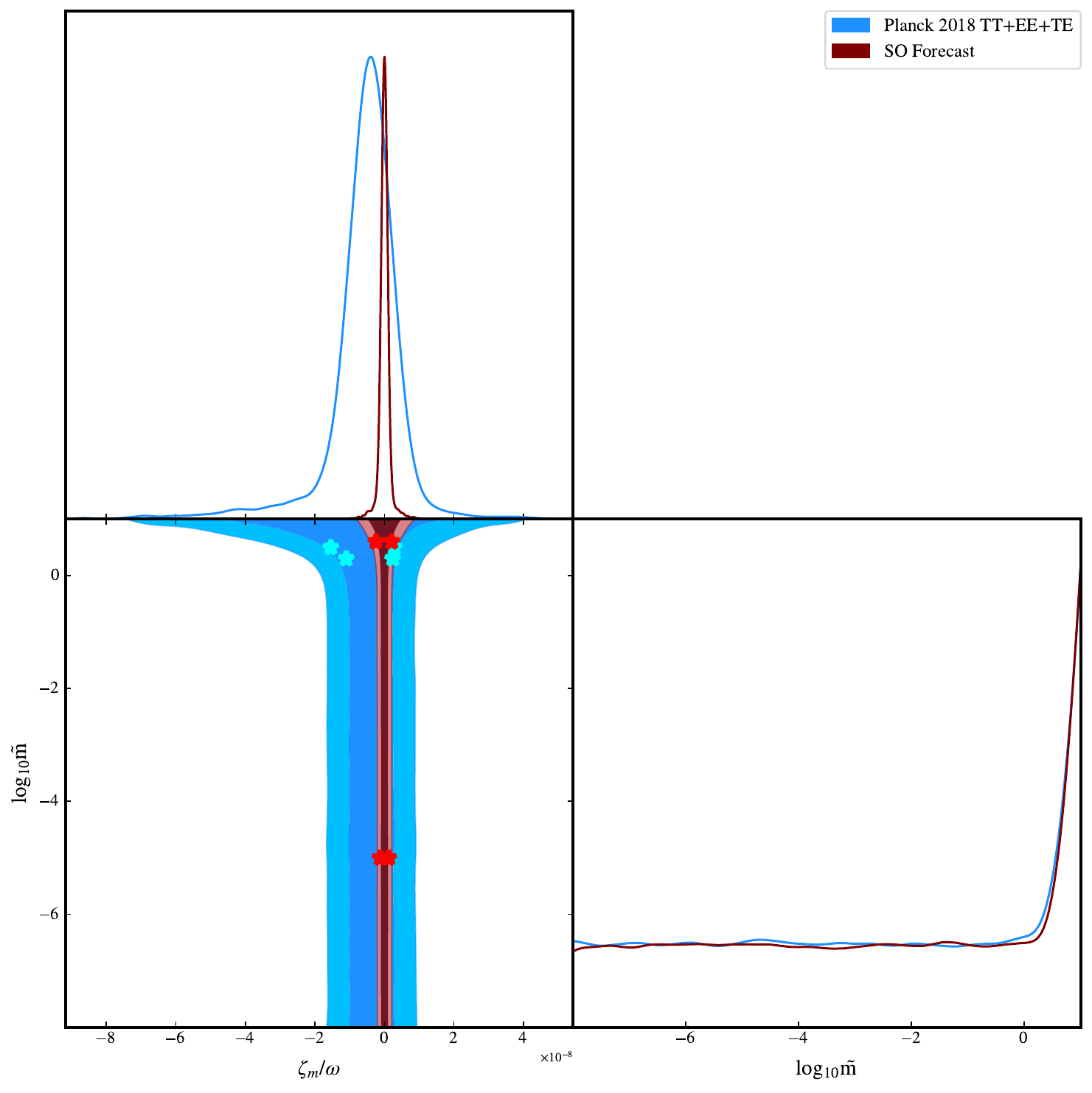}
\caption{$68.5\%$ (dark) and $95\%$ C.L.(light) contours for $\zeta_{m} / \omega$ and $\tilde{m}$ using \emph{Planck} 2018 data. Overlaid are forecasts for upcoming SO data, assuming forecast SO error bars on principal component amplitudes and a fiducial value $\zeta_m/\omega=0$. $5$-sided red (blue) stars show \emph{Planck} (SO) results from the one-dimensional MCMCs described in Sec.~\ref{sec:results}.}
\label{fig:cmb_2d_constaints}
\end{figure*}

The results are shown in Fig.~\ref{fig:cmb_2d_constaints}. As we can see, 
the constraint becomes less stringent near $\log_{10}(\tilde{m})\simeq 0$. The predicted decoupling-era $\alpha$ variation of parameter sets beyond this threshold is smaller than for lower values of $\log_{10}(\tilde{m})$. To be sure that our allowed parameter space [which is highly non-Gaussian in $\log_{10}(\tilde{m})$] is robust, we did a series of one-dimensional {MCMC runs} at a discrete grid of $m$ values near and beyond the transition point, as also done to obtain constraints to ultra-light axions (ULAs) in Ref.~\cite{Hlozek:2014lca}. 
These are presented in Table \ref{table:1d_table_planck}. 


\begin{figure*}
\includegraphics[keepaspectratio=true,width=\textwidth]{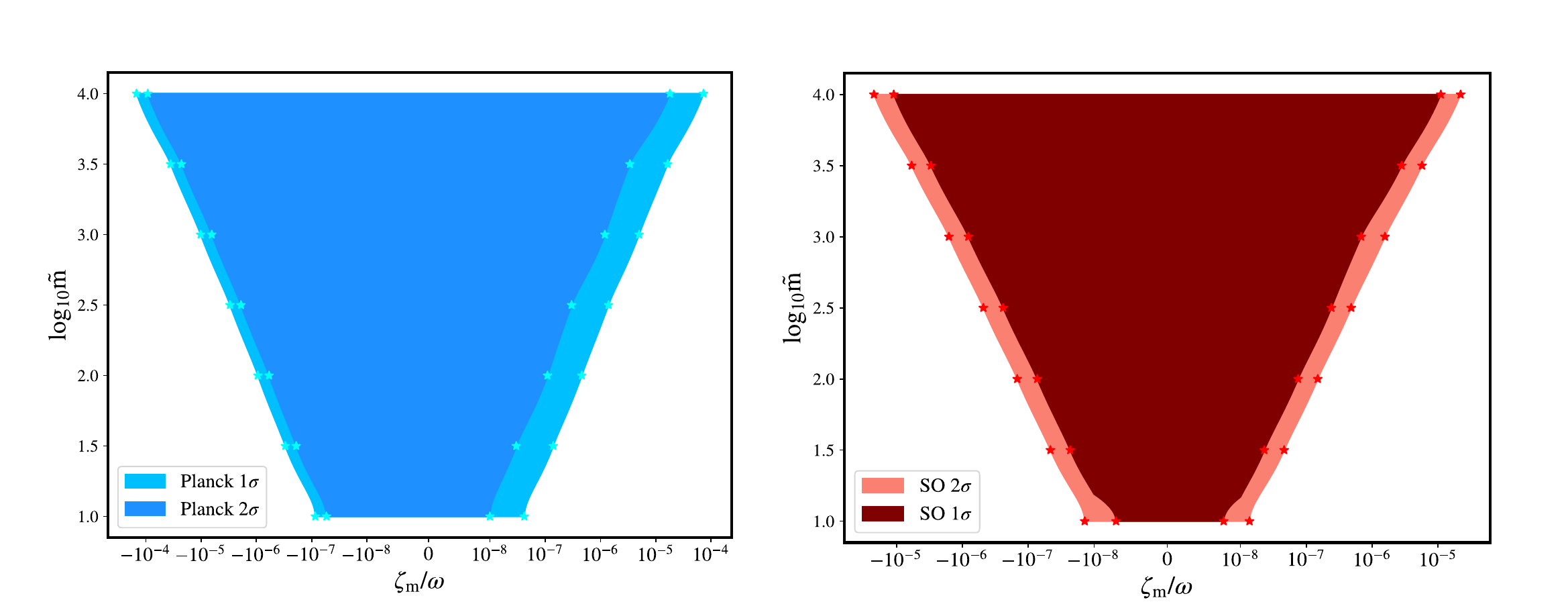}
\caption{$68.5\%$ (dark) and $95\%$ C.L.(light) contours for $\zeta_{m} / \omega$ and $\tilde{m}$ using our SO forecast and \emph{Planck} data, obtained with the adaptive solver and one-dimensional \textsc{emcee} runs. }
\label{fig:cmb_1d_planck_data_so_forecast}
\end{figure*}

\begin{figure}
\includegraphics[keepaspectratio=true,width=\columnwidth]{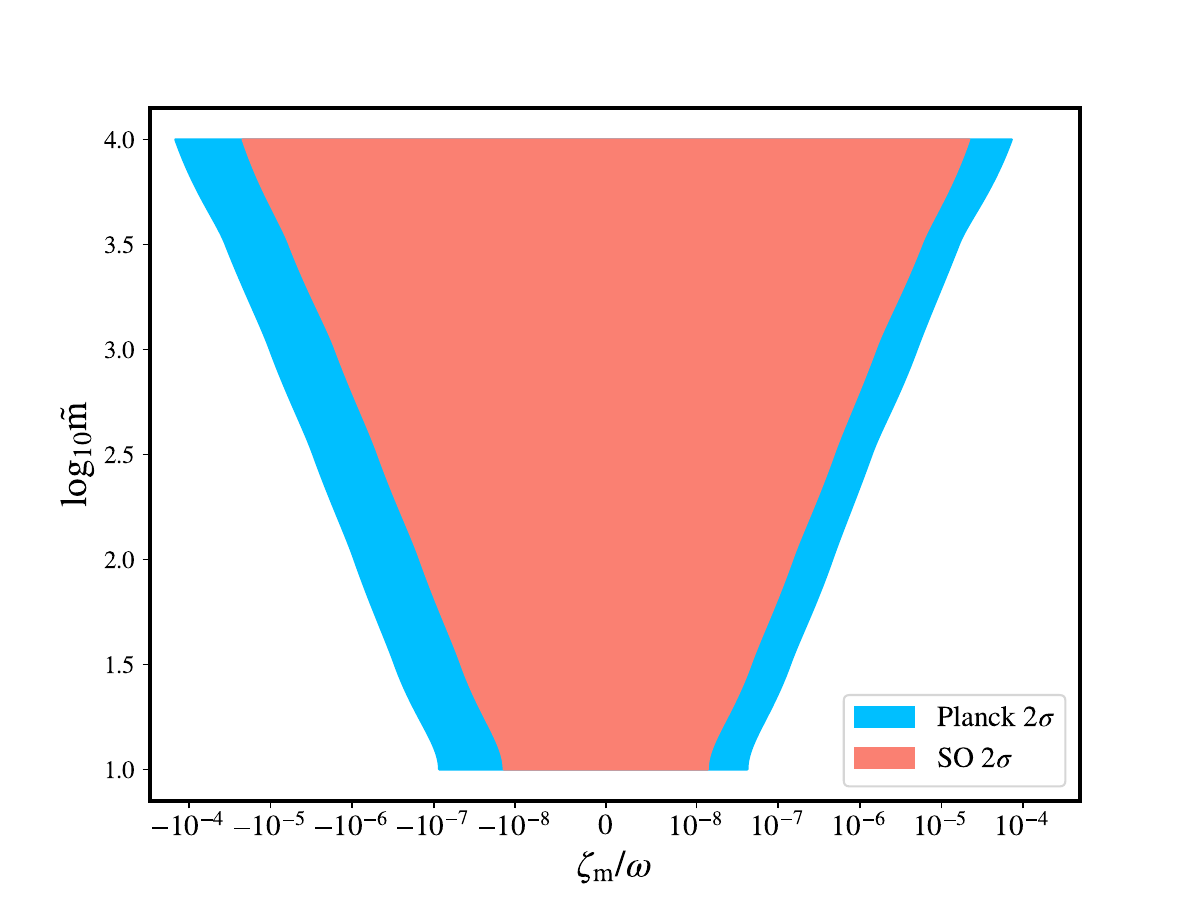}
\caption{$95\%$ C.L. (light) contours for $\zeta_{m} / \omega$ and $\tilde{m}$ using our SO  forecast and \emph{Planck} data, obtained with the adaptive solver and one-dimensional \textsc{emcee} runs. }
\label{fig:cmb_1d_so_forecast}
\end{figure}

\begin{figure*}[h]
\includegraphics[keepaspectratio=true,scale=.7]{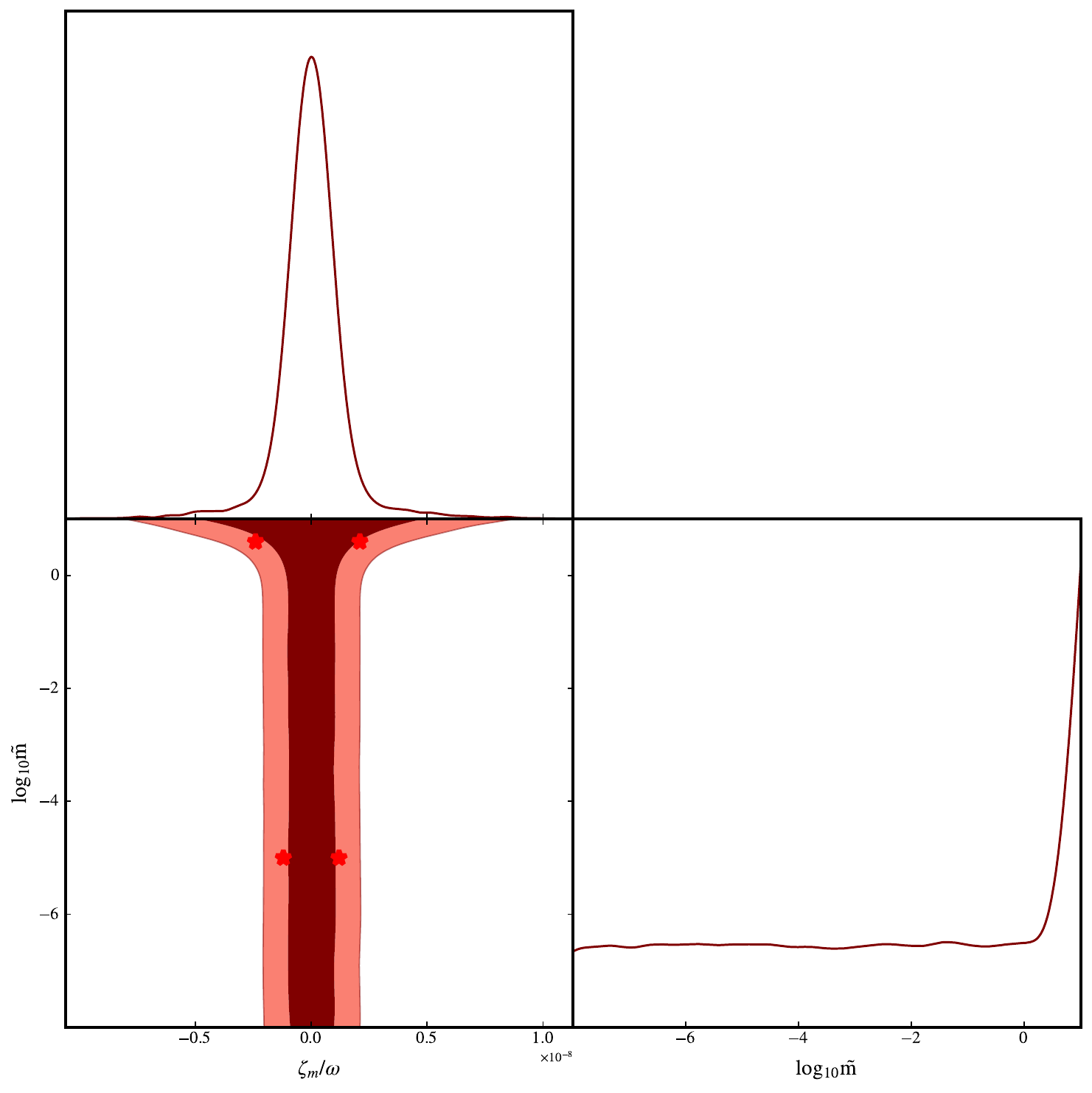}
\caption{$68.5\%$ (dark) and $95\%$ C.L.(light) contours for $\zeta_{m} / \omega$ and $\tilde{m}$ using our SO forecast. Overlaid are forecasts for upcoming SO data, assuming forecast SO error bars on principal component amplitudes and a fiducial value $\zeta_m/\omega=0$. $5$-sided stars show SO forecasts from the one-dimensional MCMCs described in Sec.~\ref{sec:results}.}
\label{fig:cmb_2d_so_forecast}
\end{figure*}

We made sure that the range used for $\log_{10}(\tilde{m})$ overlapped with that used in the two-dimensional MCMCs, finding agreement that validates both sets of constraints (the constraints from overlapping one-dimensional simulations are superimposed with stars in Fig.~\ref{fig:cmb_2d_constaints}). {This confirms that our two-dimensional constraints are not dominated by our choice of prior for $\log_{10}(\tilde{m})$. Additionally, we computed the data-theory $\chi^{2}$ on a coarse grid and obtained preliminary limits that validate our ultimate results (from both one and two-dimensional MCMCs).}

For $\log_{10}(\tilde{m})\geq 1$, we used the adaptive solver described above. Chains ($24$ for every mass) for
$\log_{10}(\tilde{m})\leq 2$ were $3000$ samples long, while those with $\log_{10}(\tilde{m})>2$ were $4500$ samples long. The autocorrelation length was $\leq 77$ samples for the high $\log_{10}(\tilde{m})$ cases, and $\leq 30$ for the low $\log_{10}(\tilde{m})$ cases, and are thus all converged. The consolidated results (constraints and forecasts) at high values of $\log_{10}(\tilde{m})$ are shown in Figs.~\ref{fig:cmb_1d_planck_data_so_forecast}-\ref{fig:cmb_1d_so_forecast}.


Summarizing our two-dimensional results, we see that at $95\%$ C.L., \emph{Planck} data impose the constraint $\zeta_{m}/\omega \leq 9.3 \times 10^{-9}$, with future SO data offering an order-magnitude improvement in sensitivity. This constraint relaxes nearly completely for $\log_{10}(\tilde{m})\geq 1.$ For SO, we find that at the lowest $\tilde{m}$ values, SO would be sensitive to values of $\zeta_{m}/\omega\simeq 2.2\times 10^{-9}$ and higher, as can be seen separately in Fig.~\ref{fig:cmb_2d_so_forecast}. These results are also shown in Table \ref{table:1d_table_so}.

It is interesting to examine the change induced in CMB observables for parameter values saturating our constraints in order to understand what features drive our sensitivity to the BSBM model. We have $\Delta C_{\ell,{\rm j}}^{\rm XY}$ for each of the PCs obtained via
\begin{align}
\frac{d C_{\ell,{\rm j}}^{\rm XY}}{d \rho _{j}}=&\sum_{i}\frac{dC_{\ell}^{\rm XY}}{dq_{i}}\frac{dq_{i}}{d\rho_{j}}\nonumber \\=&\sum_{i}\frac{dC_{\ell}^{\rm XY}}{dq_{i}}\int dz E_{j}(z)f_{i}(z),
\end{align}
where $X,Y\in\left\{{\rm TT,EE, TE}\right\}$.  We then use these expressions to calculate the BSBM-induced change to observables, applying the fact that 
\begin{equation}
\Delta C_{\ell, \mathrm{BSBM}}^{XY}=\sum_{j=1}^{\mathrm{N}_{\mathrm{pc}}} \rho_{{\rm BSBM},j} \frac{d C_{\ell,{\rm j}}^{{\rm XY}}}{d\rho_{j}}.
\end{equation}
The changes to $C_{\ell}$ for \emph{Planck} best-fit values of $\zeta_{m}/\omega$, as well as $68.5\%$ and $95\%$ C.L. constraint-saturating values, are shown in Fig.~\ref{fig:delta_cl_frac_err}, along with the same quantities for SO. For this plot we used $\log_{10}(\tilde{m})=-3.0$. The changes are normalized to the cosmic variance per multipole \begin{equation}\sigma_{C_{\ell}^{\rm XX}}=\sqrt{\frac{2}{2l+1}}C_{\ell}^{\rm XX} \end{equation} of the fiducial model. To avoid spikes near zero crossing of TE, we use the usual convention that 
\begin{equation}
\sigma_{C_{\ell}^{\rm TE}}=\sqrt{\frac{2}{2l+1}}\sqrt{\left(C_{\ell}^{\rm TE}\right)^2+C_{\ell}^{\rm TT}C_{\ell}^{\rm EE}}.
\end{equation}

We see that in temperature, the dominant effect is a decrease in high-$\ell$ anisotropies. This corresponds to a shift of the diffusion damping tail to lower-$\ell$ (larger angular scales). Larger positive values of $\zeta_{m}/\omega$ correspond to larger values of $\alpha$ in the past, more efficient scattering, later decoupling, and a surface of last-scattering closer to the observer, driving features to lower $\ell$. In temperature, this geometric effect dominates over higher scattering rates yielding lower diffusion-damping lengths (which would enhance rather than depress low-$\ell$ anisotropies).

Higher values of $\zeta_{m}/\omega$ mean larger values of $\alpha$ in the past. This means that Thomson scattering rates were higher, and $E$-mode polarization anisotropies were generated more efficiently at scales at $\ell$ below the Silk damping scale.

\begin{figure*}
    \includegraphics[width=\textwidth]{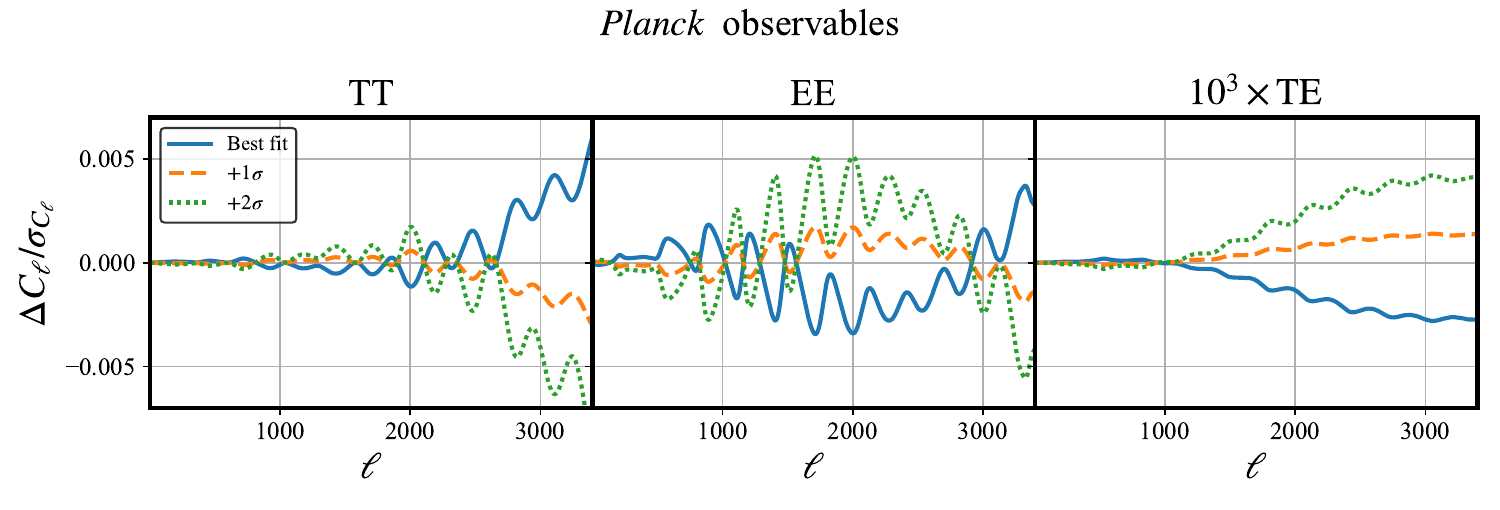}\hfill  
    \includegraphics[width=\textwidth]{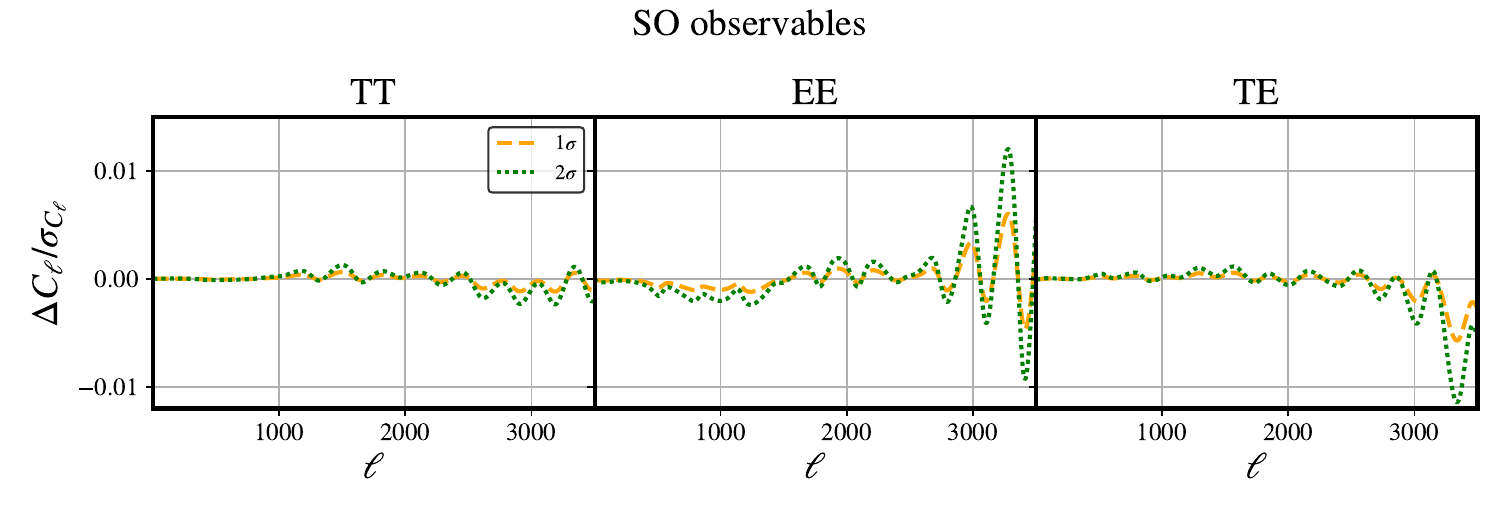}
    \caption{{Top row:} fractional change $\Delta C_{\ell}/\sigma_{C_{\ell}}$ for \emph{Planck} 2018 best fit values, and for values saturating $68.5\%$ C.L and $95\%$ C.L. constraints to $\zeta_m/\omega$, all for $\log_{10}(\tilde{m})=-3.0$. 
    Here $\sigma_{C_{\ell}}$ is the cosmic variance per multipole.
    \emph{Bottom row:} same quantity, now assuming assuming models that saturate SO error bars (themselves determined using the fiducial model).}\label{fig:delta_cl_frac_err}
\end{figure*}

\begin{figure}[h]
\includegraphics[keepaspectratio=true,scale=.4]{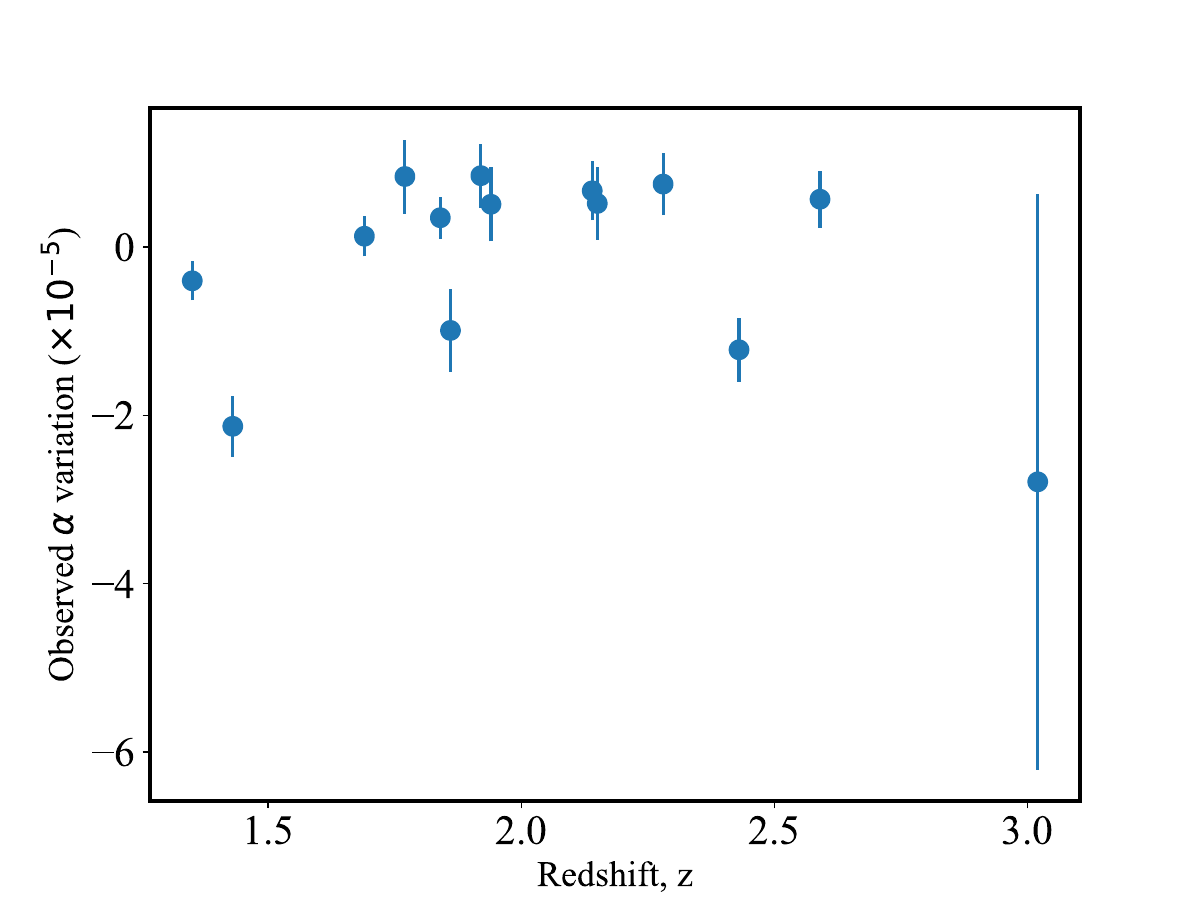}
\caption{Compilation of possible $\alpha$ variation as a function of time, inferred from analysis of QSO spectra, from Ref.~\cite{2017RPPh...80l6902M}.}
\label{qso_errorbar}
\end{figure}

\begin{table}[h]
\caption{Constraints on $\zeta_{m}/ \omega$ at different masses from \emph{Planck} 2018 analysis using one-dimensional \textsc{emcee} runs and a fixed set of $\tilde{m}$ values.}
\begin{center}
\begin{tabular}{|c|c|} 
\hline
 $\text{log}_{10}{\tilde{m}}$ & $\left(\zeta_{m} / \omega\right)$ (Planck 2018) \\  \hline
$0.3$ & $ (-4.3 \pm 6.7) \times 10^{-9} $ \\
$0.5$ & $ (-6.3 \pm 9.1) \times 10^{-9}$ \\
$0.6$ & $ (-0.8 \pm 1.1) \times 10^{-8} $ \\
$0.8$ & $ (-1.4 \pm 2.3) \times 10^{-8}$ \\
\hline\hline
$1$ & $ (-2.2 \pm 3.2) \times 10^{-8}$ \\
$1.5$ & $ (-0.8 \pm 1.1) \times 10^{-7}$ \\
$2$ & $ (-2.4 \pm 3.5) \times 10^{-7}$ \\
$2.5$ & $ (-0.8 \pm 1.1) \times 10^{-6}$ \\
$3$ & $ (-2.6 \pm 3.8) \times 10^{-6}$ \\
$3.5$ & $ (-0.96 \pm 1.3) \times 10^{-5}$ \\
$4$ & $ (-3.7 \pm 5.5) \times 10^{-5}$ \\
\hline
\end{tabular}
\end{center}
\label{table:1d_table_planck}
\end{table}

\begin{table}[h]
\caption{Forecast sensitivity levels on $\zeta_{m}/ \omega$ at different masses from the SO forecast, using one-dimensional \textsc{emcee} runs and a fixed set of $\tilde{m}$ values.}
\begin{center}
\begin{tabular}{|c|c|} 
\hline
 $\text{log}_{10}{\tilde{m}}$ &$\left(\zeta_{m} / \omega\right)$ (SO) \\  \hline
$-5 $& $ (0.0 \pm 1.2) \times 10^{-9}$ \\
$0.6 $& $ (0.2 \pm 2.1) \times 10^{-9} $ \\
$0.8 $& $ (0.0 \pm 4.9) \times 10^{-9} $ \\
$1$ & $ (0.1 \pm 6.9) \times 10^{-9} $ \\
$1.5$ & $ (0.0 \pm 2.3) \times 10^{-8}$ \\
$2$ & $ (0.1 \pm 7.4) \times 10^{-8}$ \\
$2.5$ & $ (0.0 \pm 2.4) \times 10^{-7}$ \\
$3 $& $ (-0.2 \pm 7.9) \times 10^{-7}$ \\
$3.5 $& $ (-0.1 \pm 2.9) \times 10^{-6}$  \\
$4 $& $ (-0.0 \pm 1.1) \times 10^{-5}$ \\
\hline
\end{tabular}
\end{center}
\label{table:1d_table_so}
\end{table}

\begin{table}[h]
\caption{Constraints on $\zeta_{m}/ \omega$ at different masses from QSO data, using one-dimensional \textsc{emcee} runs and a fixed set of $\tilde{m}$ values.}
\begin{center}
\begin{tabular}{|c|c|} 
\hline
 $\text{log}_{10}{\tilde{m}}$ &  $\left(\zeta_{m}/\omega\right)$ (QSO)\\  \hline
$-5$ & $ (-1.1 \pm 1.0) \times 10^{-7}$ \\
$0$& $ (-1.4 \pm 1.1) \times 10^{-7}$ \\
$1$ & $ (-1.4 \pm 1.2) \times 10^{-6} $\\
$1.5$ & $ (-1.24 \pm 0.89) \times 10^{-5} $\\
$2$ & $ (-7.4 \pm 0.83) \times 10^{-5} $\\
$2.5$ & $ (7.8 \pm 0.82) \times 10^{-4} $\\
$3$ & $ (-0.1 \pm 0.0084) $\\
$3.5$ & $ (-0.085 \pm 0.08)  $\\
$4$ & $ (-0.01 \pm 0.008)  $\\
\hline
\end{tabular}
\end{center}
\label{table:1d_table_qso}
\end{table}

\begin{figure}[h]
\includegraphics[keepaspectratio=true,scale=.3]{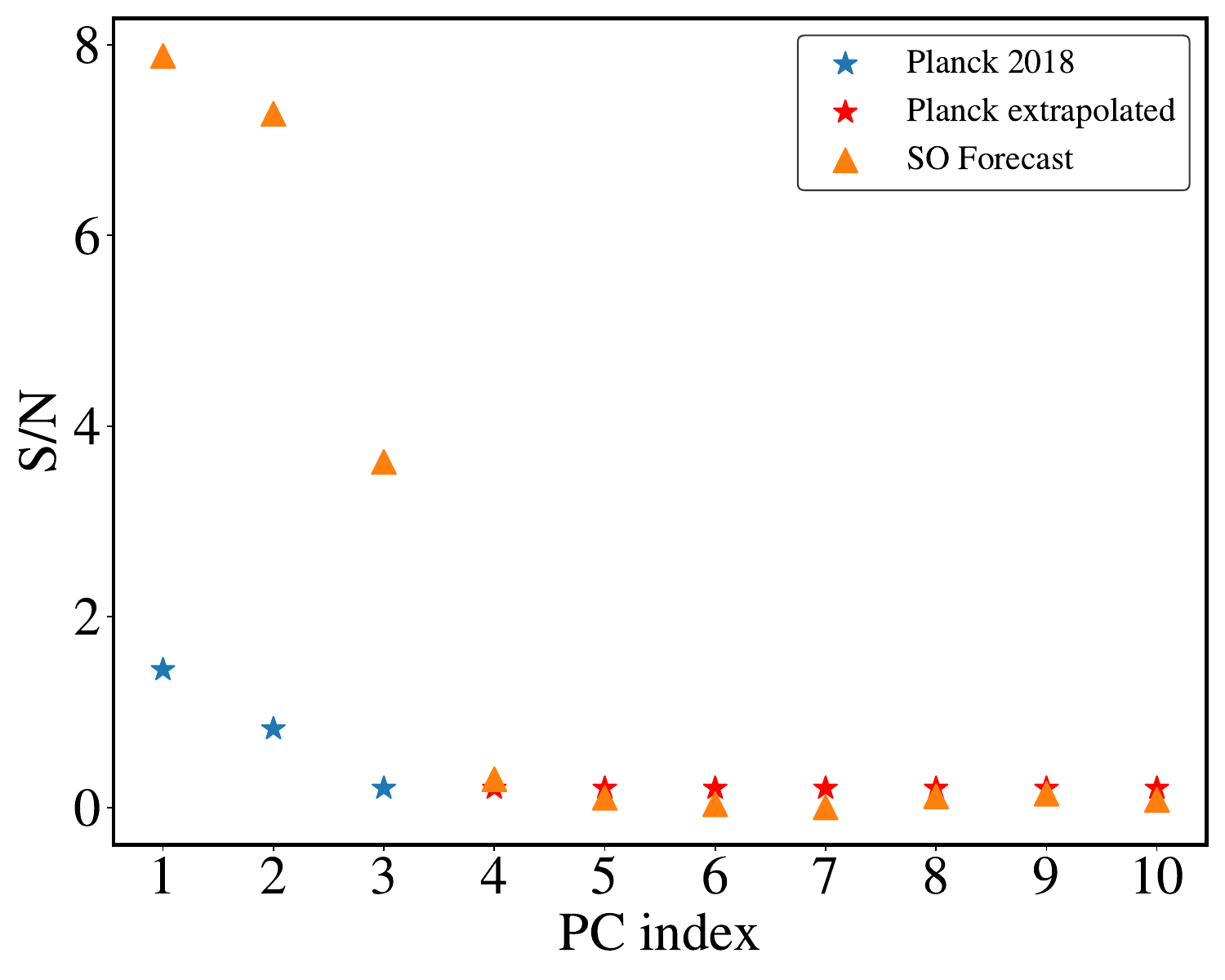}
\caption{Signal-to-noise ratio as a function of PC index, assuming true $\zeta_{m}/\omega$ that saturate \emph{Planck} $95\%$ C.L. constraints with $\tilde{m}=0$. PC eigenvalues for higher indices in \emph{Planck} are extrapolated to estimate SNR (as the eigenvalues appear to follow a clear power law) and are shown in red for clarity
(red).}
\label{fig:s_n}
\end{figure}

\begin{figure*}
\includegraphics[width=0.85\textwidth]{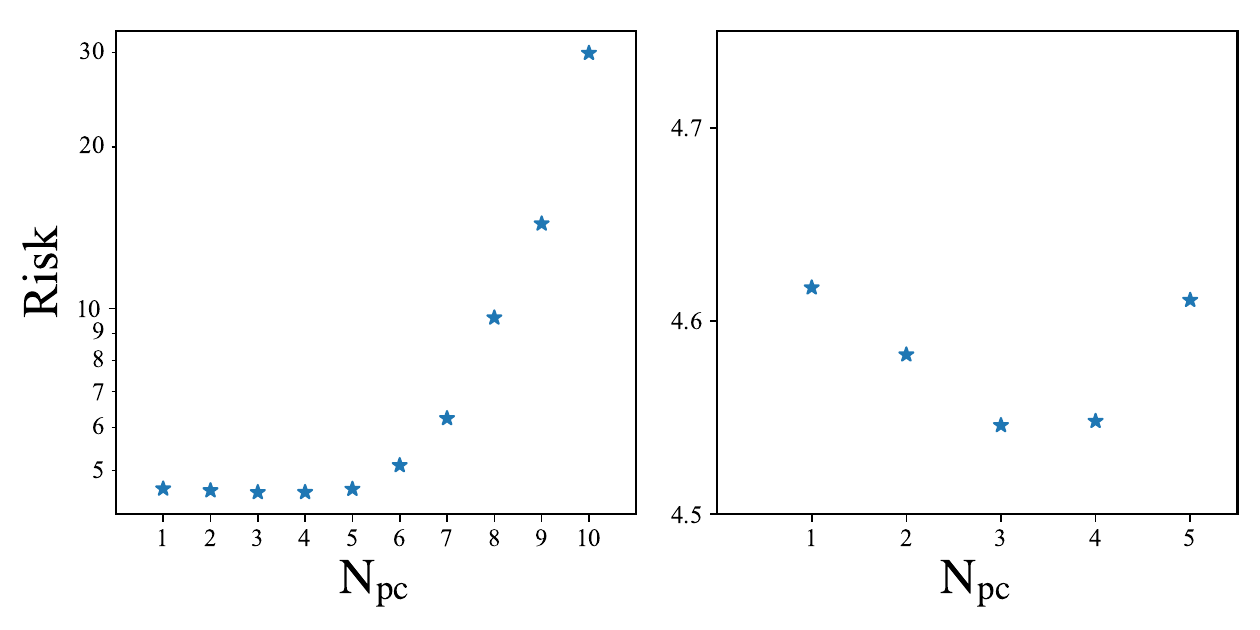}
\caption{Risk factor for the BSBM model, as a function of the number of PCs used for SO sensitivity levels, assuming $\zeta_m/\omega$ values that saturate \emph{Planck} $95\%$ C.L. constraints with $\tilde{m}=0$. \emph{Left panel:} Full range. \emph{Right panel:} Zoom in to resolve presence of a minimum in the risk.}
\label{fig:risk}
\end{figure*}

As a check, we also used \textsc{emcee} with a QSO data set and error bars shown in Fig.~\ref{qso_errorbar} to probe the BSBM parameter space. We used the $\chi^2$ between the model BSBM variation $\Delta\alpha_{\rm M}(z)$ (for any given values for $m$ and $\zeta_{m}/\omega$) and the reported variation inferred from QSO data $\Delta\alpha_{\rm D}(z)$:
\begin{equation}
    \chi^2 = \sum_{i} \frac{\left[\Delta\alpha_{\text{M}}(z_{i}) - \Delta\alpha_{\text{D}}(z_{i})\right]^2}{\sigma_i^2}.
\end{equation} 

The QSO redshifts are denoted $z_{i}$ and the measurement errors in $\Delta \alpha_D(z)$ are denoted $\sigma_{i}$. For the QSO analysis, we used $32$ one-dimensional MCMC chains of $2000$ samples each for each $\tilde{m}$ value. The autocorrelation length was $\sim 17$ samples, far less than
$2000/50=40$, and so the chains are converged.

Our reanalysis of the QSO data summarized in Ref.~\cite{2017RPPh...80l6902M} (mostly from the Keck HIRES spectrograph) yields a $95\% $ C.L. limit $\zeta_{m}/\omega\leq 2.8 \times 10^{-7}$ when $\log_{10}(\tilde{m})=-5$. Translating into the different normalization of BSBM couplings used there ($\tilde{\zeta}=8\pi \zeta_m/\omega$), this implies $\tilde{\zeta}\leq 7 \times 10^{-6}$, consistent with the limits given in Ref.~\cite{2017RPPh...80l6902M}. QSO results for higher $\tilde{m}$ are given in Table \ref{table:1d_table_qso}. The \emph{Planck} limits presented in this work (alone, without a QSO prior) to BSBM parameters are thus tighter than those imposed by the HIRES data.

Recently, extremely precise constraints to $\alpha$ variation have been obtained with the ESPRESSO spectrograph at the European Southern Observatory (ESO) and applied to test the BSBM and related models \cite{Martins:2022pue,Vacher:2023gnp}. That work 
includes an external constraint imposed from the CMB, but features QSO (and other more local) results. A $95\%$ C.L. limit of $\tilde{\zeta}\leq 1.5 \times 10^{-6}$ is given there (obtained from QSO data), equivalent to $\zeta_m/\omega\leq 6.0\times 10^{-8}$. This limit folds in a significant nonzero centroid for the PDF of $\zeta_m/\omega$ in their analysis, however, and the sensitivity of that work (with QSO data set) has a $95\%$ error bar on $\zeta_m/\omega$ of $\sim 3.3 \times 10^{-7}$, less stringent than  our \emph{Planck} analysis.

Our SO forecasts predict a sensitivity level better than this, and so it stands to reason that future CMB experimental efforts (e.g. S4, HD) will outpace some lower-$z$ probes of $\alpha$ variation in testing the BSBM model and related ideas. More broadly, this is an independent technique, depending on different physics, with different systematics, and evidence for the model at high-redshift could be reconciled with low-redshift upper limits in the context of a theory with different time evolution than initially assumed (the same could be said for a high-redshift upper limit in conflict with potential low-redshift evidence).
 
In future work, we will investigate the power of combining QSO and CMB data sets. We have developed a version of the Boltzmann code \textsc{class} which evolves the spatial fluctuations in the scalar field of the Bekenstein model. Our code includes both gravitational effects and nonminimal coupling of $\psi$ to DM. 

While we are exploring the possible impact of this \emph{spatial} fluctuation on CMB observables, we have already found that the predicted angular spectrum of $\alpha$ variations at decoupling is blue. The observable impact of these fluctuations is currently only computable within a separate universe limit (see Ref.~\cite{Heinrich:2016gqe} for a discussion) and produces a spatial variation (on scales that are super-horizon at recombination) in $\alpha$ with a root mean-squared value at the $\sim 10^{-5}$ fractional level for values $\zeta_{m}/\omega=1$. This signal is small, but the correlation with the underlying dark-matter density field (and possible early-Universe quantum contributions to $\psi$ fluctuations), could still induce an observable signal through non-Gaussian signatures in the CMB along the lines described/sought in Refs.~\cite{Sigurdson:2003pd,OBryan:2013nip,Smith:2017ndr}. 

A more detailed discussion of the results and methods of those efforts is beyond the scope of this work and will be presented in a future manuscript \cite{yuan}. The small amplitude of the signal (due to the causal growth of $\psi$ fluctuations and the smallness of primordial fluctuations) is consistent with predictions of Ref.~\cite{Barrow:2005sv}. 

To understand which PCs are driving constraints, it is interesting to examine the SNR of each mode for parameter values saturating the $95\%$ C.L. BSBM model constraints for $\tilde{m}=0$ with \emph{Planck} data, applying Eq.~(\ref{eq:permode_snr}). These quantities are shown in Fig.~\ref{fig:s_n}. We see that the first two PCs have significant constraining power for \emph{Planck}, while three PCs of SO data will be constraining. Unsurprisingly, SO will generally have a much higher SNR for the BSBM model.

We also compute the risk [Eq.~(\ref{eq:risk})] as a function of $N_{\rm PC}$, the number of PCs used in the analysis, assuming SO noise levels and a signal saturating the \emph{Planck} 95$\%$ constraint level for $\tilde{m}=0$. The results are shown in Fig.~\ref{fig:risk} for SO, indicating that three PCs should suffice to test the BSBM model adequately while minimizing risk.

\section{Conclusions}
\label{sec:conclusions}
We have used \emph{Planck} 2018 data and its principal components for variations in $\alpha$ to constrain the BSBM theory of varying $\alpha$, obtaining a $95\%$ C.L. constraint of $\zeta_{m}/\omega\leq 9.3 \times 10^{-9}$. Assuming the null hypothesis holds, we have found that the Simons Observatory will have sensitivity to values as low as $\zeta_{m}/\omega=2.2 \times 10^{-9} $. These results apply not only to the BSBM theory, but also to a related family of theoretical ideas, such as the string dilaton, the supersymmetric BSBM, the gaugino-driven modulus, and Brans-Dicke electromagnetism \cite{Olive:2001vz}.

Looking forward, it will be interesting to extend our results to further theoretical models for $\alpha $ variation [including, for example, a scalar-field potential and coupling function $\omega(\psi)$ \cite{Barrow:2008ju,Barrow:2011kr}], or models where other coupling constants (e.g. $m_{e}$ \cite{Barrow:2005qf}, $G$ \cite{Hojjati:2011xd,Hojjati:2013xqa,Koyama:2015vza,Pogosian:2016pwr,Espejo:2018hxa}, or even $c$ \cite{Albrecht:1998ir,Magueijo:2000zt,Magueijo:2003gj}) are dynamical.

Using kernel-density estimates of the full-PCA +$\Lambda$CDM likelihood, we will more fully probe the covariance of BSBM model parameters with standard cosmological parameters. Causality dictates that variations of $\alpha$ (or other fundamental parameters) in time require variations of $\alpha$. While these variations are likely to be very small, they would induce non-Gaussian signatures in the CMB \cite{Sigurdson:2003pd,OBryan:2013nip,Smith:2018rnu}, and it would be useful to explore if this signal could be better extracted by applying CMB delensing, and harnessing the cross-correlation of $\alpha$ with the underlying density field \cite{He:2015msa}.  The induced CMB bi- and trispectra by these models have, in principle, a distinct shape from more standard effects like CMB lensing \cite{He:2015msa,Heinrich:2016gqe}. In future work, we will assess if this can be used to better extract spatial variations in $\alpha$ from the CMB.

Thinking towards future measurements of CMB anisotropies, it will be interesting to explore the power of the planned CMB-S4 experiment \cite{Abazajian:2016yjj,Abazajian:2019eic}, as well as more futuristic concepts like CMB-HD \cite{Sehgal:2019ewc}, which could probe the CMB at much smaller angular scales than ever before, promising much improved leverage on varying fundamental constants.

In coming years, new cosmological frontiers will open, with likely measurements of the neutral $21$-cm signature of cosmic reionization and the cosmic dark ages \cite{Pritchard:2011xb}. The global 21-cm signal and anisotropies should be strongly sensitive to $\alpha$ \cite{Khatri:2007yv,Lopez-Honorez:2020lno}. Furthermore, CMB spectral distortions from the Silk damping and recombination eras would have spatial and frequency dependence that is strongly dependent on $\alpha$ \cite{Hart:2022agu}. In future work, we will assess the sensitivity of these powerful measurements to the full family of theories enumerated here. The era of precision cosmology is here, and we should look forward to harnessing its data products, not only to characterize the energy budget of the Universe but also to test the constancy of the fundamental parameters. 

\acknowledgments 
The authors thank Vivian Miranda and Jeremy Sakstein for helpful conversations. We thank Tristan Smith for many helpful discussions, especially on the issue of energy conservation in the BSBM theory. {We thank an anonymous referee for questions/comments which helped improve the quality of this paper.} H.~T. was supported by a Velay Fellowship and travel funding from the Marian E. Koshland Integrated Natural Sciences Center (KINSC) at Haverford College. D.~G., H.~T., M.~B., and J.~Cr. were supported by the U.S. National Science Foundation through grant award NSF 2112846. E.~B. was supported by Haverford College's KINSC Summer Scholars fellowship. {L.H. and J.~Ch. were supported by the ERC Consolidator Grant {\it CMBSPEC} (No.~725456). J.~Ch. was furthermore supported by the Royal Society as a Royal Society University Research Fellow at the University of Manchester, UK (Grant No.~URF/R/191023).} This work used the Hannah Computing Cluster, which is run by Haverford College, the Strelka Computing Cluster, which is run by Swarthmore College, and the High Performance Computer Cluster which is run by the University of California, Riverside. We thank Joe Cammisa for support with Hannah. We thank Jason Simms and Andrew Reuther for support with Strelka. The land on which Haverford College stands is part of the ancient homeland and unceded traditional territory of the Lenape people. We pay respect to Lenape peoples, past, present, and future, and their continuing presence in the homeland and throughout the Lenape diaspora. We acknowledge the authors of the \textsc{numpy} \cite{harris2020array} and \textsc{scipy} \cite{2020SciPy-NMeth} libraries. We thank the authors of the \textsc{emcee} software used in our MCMCs \cite{Foreman-Mackey:2012any}, and the \textsc{GetDist} package used for visualizing the MCMC results \cite{Lewis:2019xzd}. 

\begin{appendix}
\section{Electromagnetic sourcing of $\alpha$ variation}
\label{sec:zeta}
\subsection{Argument for shielding of electrostatic energy}
Here we recapitulate Ref.~\cite{Bekenstein:2002wz}'s argument that electrostatic contributions to $\zeta_{m}$ are shielded. We follow the discussion there but provide additional details as needed. We begin with the static contributions to the Poisson equation:
 \begin{equation}
     \label{eq:psi_eom} \nabla^2\psi=4\pi\kappa^2\left[\sum_i{\frac{\partial m_ic^2}{\partial\psi}\delta^3(\mathbf{x}-\mathbf{z}_i)}+\frac{1}{4\pi}e^{-2\psi}{E}^2\right].
 \end{equation} The sum here is over all source particles and the constant $\kappa=l/\sqrt{4\pi\hbar c}$ is a renormalized Brans-Dicke (kinetic) coupling for $\psi$.
 
 Bekenstein integrates the total contributions from each particle in a volume $\mathcal{V}$ in the first term of the rhs of Eq.~(\ref{eq:psi_eom}) and then replaces $(\partial m_i c^2)/\partial \psi$ with $\kappa^{-1}e_{0i}\tan[\kappa\Phi(\mathbf{z}_i)]$ (Eq.~(43) of Ref.~\cite{Bekenstein:2002wz}). 
 
 The first term [on the rhs of Eq.~(\ref{eq:psi_eom})] then becomes 
 \begin{align*}
     &\sum_i{\frac{\partial m_ic^2}{\partial\psi}\delta^3(\mathbf{x}-\mathbf{z}_i)} \\
     &\rightarrow -\frac{1}{\mathcal{V}}\int_\mathcal{V}d^3x\sum_{i\in\mathcal{V}}\kappa^{-1}e_{0i}\tan[\kappa\Phi(\mathbf{z}_i)]\delta^3(\mathbf{x}-\mathbf{z}_i) \\
     &= -\frac{1}{\mathcal{V}}\sum_{i\in\mathcal{V}}e_{0i}\tan[\kappa\Phi(\mathbf{z}_i)].
 \end{align*}
Taylor expanding $\tan[\kappa\Phi(\mathbf{z}_i)]$ about $\Phi(\mathbf{z_i})$ and discarding terms of $\mathcal{O}(\kappa^3\Phi^3)$, the expression 
 \begin{equation}
    \label{eq:term1_approx}
     -\frac{1}{\mathcal{V}}\sum_{i\in\mathcal{V}}e_{0i}\Phi(\mathbf{z}_i).
 \end{equation} is then obtained for the first term.
 
The second term is somewhat more complicated. Modified Maxwell equations of the BSBM theory imply that $\mathbf{E}$ with $\mathbf{E}=-e^{2\psi}\boldsymbol{\nabla}\Phi$ (Eq.~(41) of Ref. \cite{Bekenstein:2002wz}), and since $e^{2\psi}=\sec^2{\kappa\Phi}$ (Eq.~(45) of Ref. \cite{Bekenstein:2002wz}), linearization yields
 \begin{equation}
     e^{2\psi}=\sec^2(\kappa\Phi)\approx 1 + \kappa^2\Phi^2 + \mathcal{O}(\kappa^4\Phi^4).
 \end{equation}
 Discarding terms of $\mathcal{O}(\kappa^4\Phi^4)$, the spatial average of the second term is
 \begin{equation}
    \label{eq:full_E_spatial_average}
     \frac{1}{4\pi}e^{-2\psi}\mathbf{E}^2\rightarrow\frac{1}{4\pi\mathcal{V}}\int_\mathcal{V}d^3x[(\boldsymbol{\nabla}\Phi)^2+\kappa^2\Phi^2\mathbf{E}^2].
 \end{equation}

Integrating the first term of this integral using the divergence theorem and applying Eq.~(\ref{eq:psi_lap}), one finds
 \begin{equation}
    \label{eq:psi_lap}    \nabla^2\Phi=-4\pi\sum_ie_{0i}\delta^3(\mathbf{x}-\mathbf{z}_i).
 \end{equation}
The final approximation (applying the immediately preceding result) for the second term on the rhs of Eq.~(\ref{eq:psi_eom}) is \begin{widetext}
\begin{equation}
    \label{eq:term2_approx}
     \frac{1}{\mathcal{V}}\sum_{i\in \mathcal{V}}e_{0i}\Phi(\mathbf{z}_i)-\frac{1}{4\pi\mathcal{V}}\left[\oint_{\partial\mathcal{V}}\Phi\mathbf{E}\cdot d\mathbf{s}-\kappa^2\int_\mathcal{V}d^3x\Phi^2\mathbf{E}^2\right].
 \end{equation}
 \end{widetext}
 
 Summing Eqs.~(\ref{eq:term1_approx}) and (\ref{eq:term2_approx}), it is immediately clear that the only terms remaining are the two integrals in Eq.~(\ref{eq:term2_approx}), and Ref.~\cite{Bekenstein:2002wz} argues they both are negligible--the argument goes as follows. One can approximate the surface integral as $\mathcal{V}^{-1}\langle\Phi\rangle\sum_{i\in\mathcal{V}}e_{0i}$, using Gauss's law where $\langle\Phi\rangle$ is the surface average of $\Phi$. This quantity is much less than $\mathcal{V}^{-1}\sum_{i\in\mathcal{V}}e_{0i}\Phi(\mathbf{z}_i)$, so it is negligible. 
 
The only remaining quantity is the second integral in Eq.~(\ref{eq:term2_approx}). An upper bound on $|\Phi|$ for a unit charge is obtained divided by the smallest length scale $\Phi$ varies over, which is at most $10^{-17}$\unit{\cm}, if quarks are the smallest particles of interest. Then, $\kappa^2\Phi^2\mathbf{E}^2/4\pi$ has an upper bound of roughly $10^{-34}(l/l_p)^2\mathbf{E}^2/4\pi$, where $l$ is the characteristic length  of Bekenstein's theory. This implies that using $\mathbf{E}^{2}/4\pi$ vastly overestimates electrostatic energy as a source for $\alpha$ variation in the BSBM model. 
 
\subsection{Testing electrostatic cancellation}
In order to the claim that $\psi$ arranges itself to self-shield the electrostatic contribution to its equation of motion, we conduct a particle-in-cell (PIC) plasma simulation of the early Universe (proton and neutron) plasma in \textsc{python}. These particles interact and cluster in ways that produce electric and magnetic fields. 

We compute these and use them to estimate $\zeta_m=\langle \mathbf{E}^2 - \mathbf{B}^2 \rangle /\rho_m$. We determined field quantities by solving the modified Maxwell's equations using a Fourier spectral method. The equations we solved were the following, given in both real and $k$ space:
\begin{align}
    \mathbf{E} = \mathbf{\nabla}\Phi &\Longrightarrow \tilde{\mathbf{E}}=i\tilde{\Phi}\mathbf{k}  \label{eq:normal_E} \\
    \nabla^2\Phi - -4\pi\rho &\Longrightarrow \tilde{\Phi}=-4\pi\tilde{\rho}K^2 \label{eq:Phi} \\
    \nabla^2\mathbf{A}=\frac{4\pi}{c}\mathbf{J} &\Longrightarrow \tilde{\mathbf{A}}=\frac{4\pi}{c}\tilde{\mathbf{J}}K^2
    \label{eq:B}
\end{align}
where $\mathbf{\nabla}\times\mathbf{A}=\mathbf{B}$, $\mathbf{k}=\left\{dx^{-1}\sin(k_xdx),dy^{-1}\sin(k_ydy),dz^{-1}\sin(k_zdz) \right\}$, and $K=K_x^2+K_y^2+K_z^2$ with $K_x=2dx^{-1}\sin(k_xdx/2)$ and analogous definitions for $K_y$ and $K_z$ \cite{Birdsall_1991}. These definitions for $k$ and $K_x,K_y,K_z$ come from taking the Fourier transform of the finite difference forms of $\nabla$ and $\nabla^2$ respectively. 

We use these algebraic equations to solve for the fields of interest in the simulation. At each time step, we interpolate the particles onto a spatially uniform grid using a linear spline and periodic boundary conditions. Then, we use the node locations and appropriate weighting to determine the particle number density and current density at each node. We then take the discrete Fourier transform of these quantities and solve for $\mathbf{\tilde{E}}$ and $\mathbf{\tilde{A}}$ using Eqs. (\ref{eq:normal_E}) through (\ref{eq:B}). Next, we take the inverse Fourier transform of $\mathbf{\tilde{E}}$ and $\mathbf{\tilde{A}}$ to get $\mathbf{E}$ and $\mathbf{A}$. Finally, we calculate $\nabla \times \mathbf{A}=\mathbf{B}$. $\mathbf{E}$ and $\mathbf{B}$ are then reinterpolated onto particle positions. These are the quantities used to push the particles using the relativistic Boris method \cite{2018ApJS..235...21R}. In order to calculate the discrete Fourier transforms, we used a built-in Fourier transform package from the \textsc{numpy} module for \textsc{Python} \cite{harris2020array}. 

The simulation uses $4096$ particles on a $128\times128\times128$ grid at $z=1100$. The particles are arranged uniformly on the grid initially and their initial velocities are drawn from a Maxwell-Boltzmann distribution. The domain of the simulation was a box where each side was $Nn_0^{-1/3}$ long, where $N$ is the number of particles along one side and $n_0$ is the number density of protons in the Universe at $z=1100$.  Also, $dt=0.1dx/v_0$, where $v_0$ is the mode of the initial velocity distribution for electrons. This was set to ensure that no particle traverses more than one grid cell in a time step. 

We tested the code using several methods. First, we tested the particle-pushing method by placing a single electron under the influence of a uniform electric and uniform magnetic field. In both cases, the error of our particle pusher method is of order machine precision (e.g. Fig.~\ref{fig:B_pusher_test}).

\begin{figure}[h]
    \centering
    \includegraphics{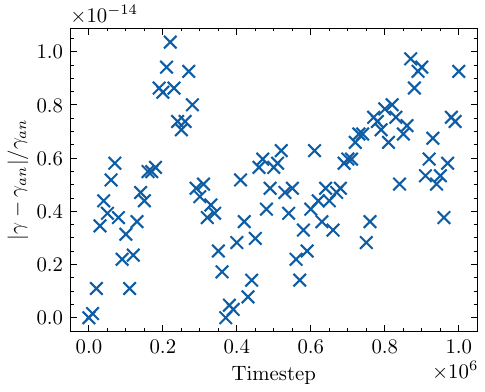}
    \caption{{The fractional error in the difference between the calculated Lorentz factor $\gamma$ and the analytic Lorentz factor $\gamma_{an}=\sqrt{1+v_{0y}^2}$} of a particle with $m=1$, $q=1$, and $c=1$ in a constant $\mathbf{B}$ field. The time step was $dt=2\pi/(10^8)$ and proceeded for $10^6$ time steps. Initial $v_0=(0,1-5\times10^{-13},0)$ and $\mathbf{B}=(0,0,10^6)$. These quantities follow the recommendations in Ref. \cite{2018ApJS..235...21R}.}
    \label{fig:B_pusher_test}
\end{figure}

Next, we tested our Fourier equation solving method by attempting to reproduce the analytical results for known functions. For example, instead of using the actual current density from the simulation, we used a known function [$f(x,y)=\sin{x}+\sin{y}$, for example] and applied the same field solver to it. This test confirmed that the Fourier solving method accurately calculates $\nabla{f}$ and $\nabla^2f$. For an additional test of the Fourier solving method, we tested it alongside a more traditional grid-based solver in a 1D plasma simulation and another 1D simulation using the same initial conditions as our final 3D simulation \cite{Mocz_2022}. We did the calculations separately from start to finish in both MKS (meters-kilograms-seconds) and CGS (Gaussian or centimeters-grams-seconds) units. The simulation produced identical results for all quantities in each context. {Additionally, we used a classic 1D two-stream instability test with two streams of particles each with $q/m=-1$ with opposite velocities. This test showed the characteristic swirls in phase space \cite{Birdsall_1991}. As a final test, we confined 2 identical particles to a line and verified that they exhibited simple harmonic motion, as is expected given the periodic boundary conditions of the simulation.}

We tested for numerical convergence. For a fixed number of particles along one axis $N$, $\mathbf{E}$ and $\mathbf{B}$ converge as the number of cells along one axis $N_x$ increases. Increasing $N_x$ increases the grid resolution, and so we would expect both $\mathbf{E}$ and $\mathbf{B}$ to converge to some value. This is indeed what we observe. As an important note, one would not expect $\mathbf{E}$ and $\mathbf{B}$ to converge as $N$ increases. This is because $n_0$ is physically set by the physics of the Universe at a given $z$. Therefore, increasing $N$ simply increases the domain size of the simulation without altering the electrodynamics of the plasma. This means that for a fixed ratio $N/N_x$, the simulation should produce identical results, which we also observed. Of course, $N$ should be set high enough to allow for sufficient interactions between particles, but once $N\approx 16$, any additional increase is not necessary.
 
Increasing $N$ higher than necessary also means that a higher $N_x$ is needed to increase the accuracy of the results. Finally, it takes about 100 time steps for the the simulation to properly equilibrate, so results to this point should be discarded. After this point however, the values of $\mathbf{E}$ and $\mathbf{B}$ converge as $N_x$ increases. 

We used the uninterpolated $\mathbf{E}$ and $\mathbf{B}$ to calculate $\zeta_m=\langle {E}^2 - {B}^2\rangle / \rho_m$. The averaging was calculated by taking the arithmetic average of the values of ${E}^2$ and ${B}^2$ at each point on the grid. Using this simulation, we estimate $\zeta_m\approx10^{-13}$ which is eight orders of magnitude lower in absolute than other estimates--this is not surprising, our calculation is obtained for a diffuse plasma on scales well above nuclear length scales--our simulation is not appropriate to test the absolute scale of $\zeta_{m}$ from nuclei in the early Universe.

It does give us the tools to test the claim that the electrostatic contribution to $\zeta_{m}$ cancels out in a specific environment that we can directly simulate. Our conclusions may need to be revised to properly account for nuclear scales, but offer an interesting first test of the (analytic) claims of electrostatic shielding made in Ref.~\cite{Bekenstein:2002wz}. 

In contrast to that approach, we calculated all the electric and magnetic fields generated through inhomogeneities in a neutral plasma like the early universe. We did not consider nucleons or macroscopic objects as sources for fields. We calculated $\mathbf{E}$ and $\mathbf{B}$  directly in the simulation using Eqs.~(\ref{eq:normal_E})-(\ref{eq:B}). However, this simulation differed from that used to estimate $\zeta_m$ because we introduced $\psi$ as a scalar field responsible for $\alpha$ variation. This required us to alter our equation for $\mathbf{E}$:
\begin{equation}
    \label{eq:Bekenstein_E}
    \mathbf{E}=-e^{2\psi}\nabla{\Phi}
\end{equation}
Because $e^{\psi}=\sec(\kappa\Phi)$, we were also able to calculate the new $\mathbf{E}$ directly in the simulation. $\mathbf{B}$ is unchanged according to Ref. \cite{Bekenstein:2002wz}. Additionally, because $\kappa$ is a free parameter, we estimate it as $8.11\times10^{-26}\: \text{cm}^{1/2}\: \text{erg}^{-1/2}$ \cite{Bekenstein:2002wz}. While this has an effect on the precise values of the terms in Eq. (\ref{eq:psi_eom}), it does not affect cancellation.  

We thus calculated all field quantities in Eq.~(\ref{eq:psi_eom}) directly. The only approximation made was in the delta function in Eq.~(\ref{eq:psi_eom}):
\begin{equation}
    \delta^3(\mathbf{x}-\mathbf{z}_i)\approx \frac{1}{a\sqrt{\pi}}\exp{\left(-\frac{(\mathbf{x}-\mathbf{z}_i)^2}{a^2}\right)}.
\end{equation} in order to properly account for finite grid resolution. The other parameters of the simulation were the same as in our simulation to estimate $\zeta_m$.

Fig.~\ref{fig:psieom} shows our results for our test of Bekenstein’s cancellation theorem. The combined rhs is only marginally different from the ${E}^2$ term. On average, the ${E}^2$ term is about {$1\times10^{-63}\: \text{erg cm}^{-3}$}, as is the combined rhs. However the $\partial m_i/\partial\psi$ term is about {$-2\times10^{-67}\: \text{erg cm}^{-3}$, about 4 orders of magnitude less than the ${E}^2$ term.} Because this term is so much smaller, it is not plotted in Fig.~\ref{fig:psieom}. Because of the large difference between the ${E}^2$ term and the $\partial m_i/\partial\psi$ term, our results do not support cancellation. 

In order to further test the cancellation theorem, we further simulated a neutral plasma at a nuclear density $n_0 = 1.2\times10^{38}$ protons/cm$^3$. Here, our simulation still does not support cancellation. In this case, the two terms in Eq.~(\ref{eq:psi_eom}) are {24 orders of magnitude apart. The ${E}^2$ term is about $10^{-16}$ erg cm$^{-3}$ and the $dm/d\psi$ term is about $10^{-40}$ erg cm$^{-3}$. }

Finally, we repeated these calculations for a proton-only plasma at the same density as the recombination-density neutral plasma. Bekenstein's arguments center on the properties of charged macroscopic objects, so it is possible that for a plasma with a net charge, the cancellation theorem holds. First, we recalculate $\zeta_m$. For the positive plasma with a $128\times 128\times 128$ grid, $\zeta_m\approx 10^{-14}$, which is slightly lower than for a neutral plasma. The results for cancellation are in Fig.~\ref{fig:pos_plasma_eom}. Here, the two terms from Eq.~(\ref{eq:psi_eom}) still differ significantly. {The $dm/d\psi$ term is about $-10^{-66}$ erg cm$^{-3}$ and the ${E}^2$ term is about $10^{-63}$ erg cm$^{-3}$.} While the two terms still differ greatly, Fig.~\ref{fig:pos_plasma_eom} demonstrates that cancellation is more plausible in this case, as there is some difference between each histogram, as we would expect if cancellation is occurring. 

 \begin{figure}[h]
    \centering 
    \includegraphics{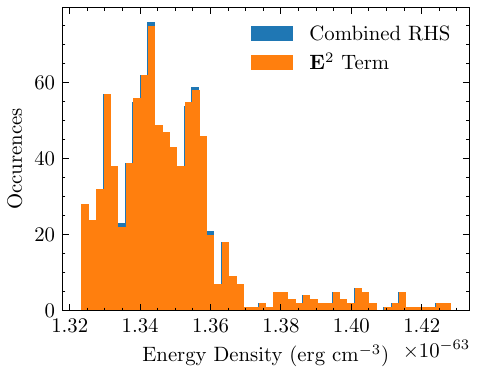}
    \caption{The values of the ${E}^2$ term and combined rhs in the equation of motion for $\psi$ for a recombination density neutral plasma. The $\partial m_i/\partial \psi$ term is not plotted because it is {4} orders of magnitude less than the other terms. Simulation used 4096 particles on a $128\times128\times128$ grid over 1000 time steps. The first 100 time steps are not plotted.}
    \label{fig:psieom}
\end{figure}

One possible reason for our observed lack of cancellation is Bekenstein's reliance on approximations to convert between microscopic and macroscopic scales. We considered instead fields calculated directly in the simulation. Some of Bekenstein's approximations to perform this conversion are robust. We tested the approximation of the second term of Eq.~(\ref{eq:psi_eom}), which is given by Eq.~(\ref{eq:term2_approx}). In the recombination-density {and nuclear density} simulations, the approximation is off by at most an order of magnitude, indicating the approximation is reasonably robust. In turn, this suggests that the problem lies with the $dm/d\psi$ term in Eq.~(\ref{eq:psi_eom}). This approximation relies more on the analytic specifics of Bekenstein's theory and is therefore more difficult to test computationally. Further work should be aimed at testing the robustness of the analytic derivations of this term. 

\begin{figure}
    \centering
    \includegraphics{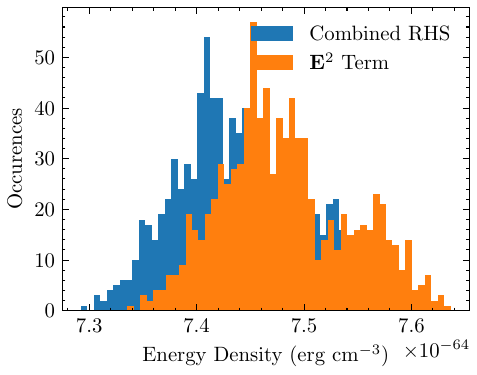}
    \caption{The values of the ${E}^2$ term and combined rhs in the equation of motion for $\psi$ in a recombination density positive plasma. The $\partial m_i/\partial \psi$ term is not plotted because it is {2} orders of magnitude less than the other terms. Simulation used 4096 particles on a $128\times128\times128$ grid over 1000 time steps. The first 100 time steps are not plotted. }
    \label{fig:pos_plasma_eom}
\end{figure}

\section{Energy Conservation}
\begin{figure*}
\includegraphics[keepaspectratio=true,scale=.35]{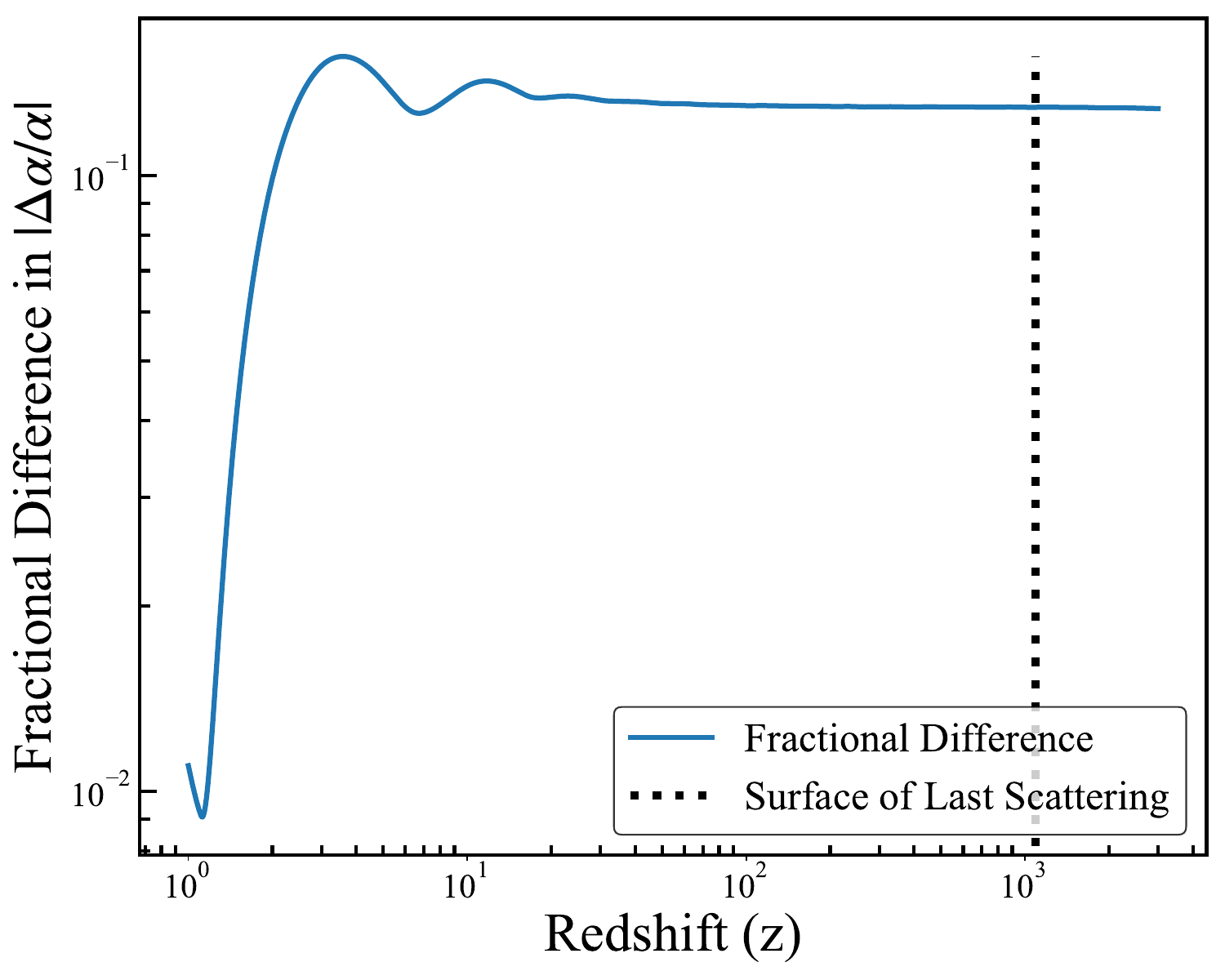}
\includegraphics[keepaspectratio=true,scale=.35]{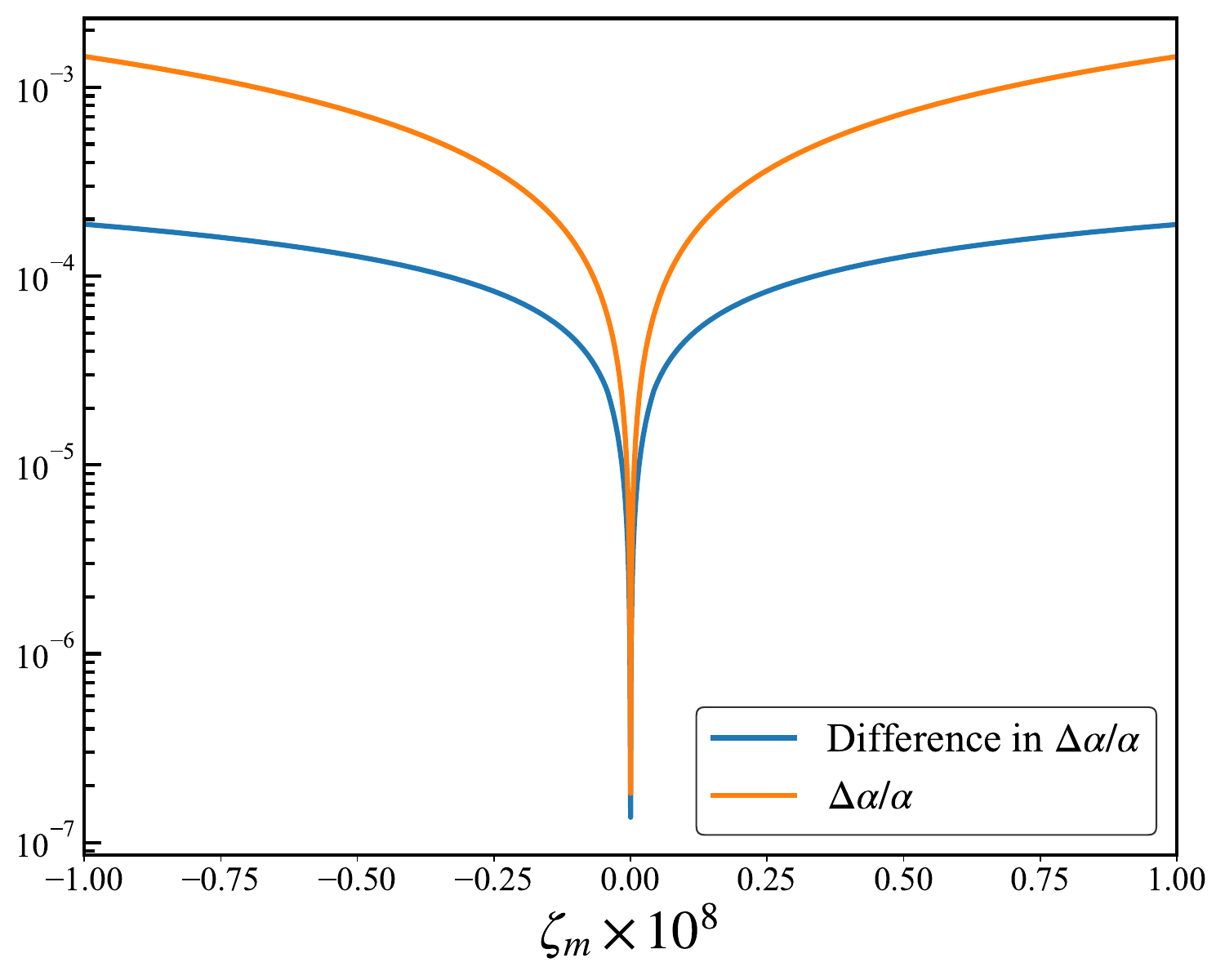}
\caption{{Left panel:} fractional error as a function of $z$ between energy-conserving and nonconserving BSBM model implementations with overall level of fine-structure constant variation, $\Delta \alpha/\alpha$.
The parameter $\zeta_m=10^{-8}$, while $\omega= 1$, consistent with the \emph{Planck} constraint. {Right panel:} Comparison of error at recombination between energy-conserving and nonconserving BSBM model implementations with overall level of fine-structure constant variation, $\Delta \alpha/\alpha$.
The parameter $\zeta_m$ is varied in the range  $-10^{-8}$ to $10^{-8}$, while $\omega= 1$.}
\label{fig:econserve_redshift}
\end{figure*}
\begin{figure*}
\includegraphics[keepaspectratio=true,scale=.35]{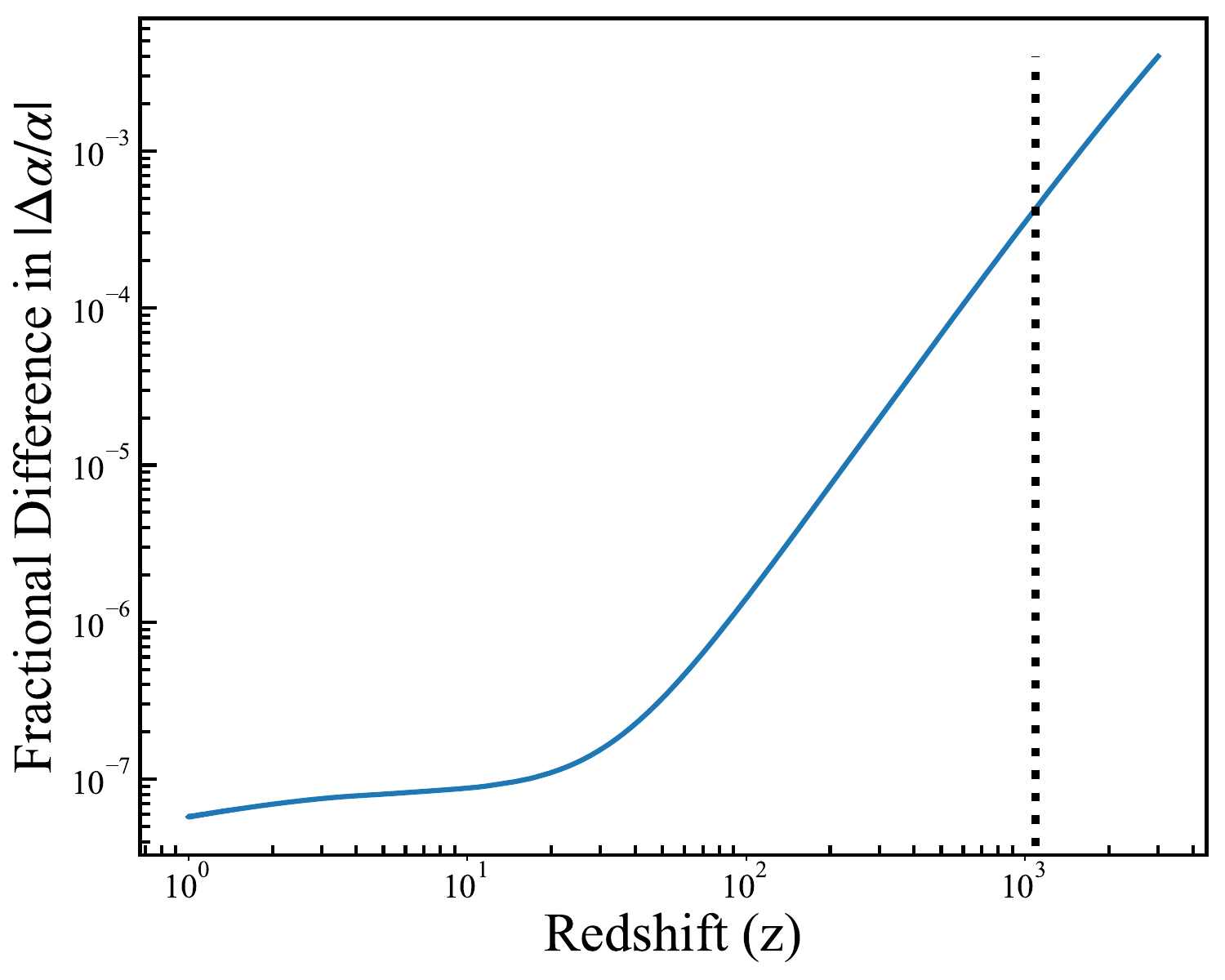}
\caption{Fractional difference in $|\Delta \alpha/\alpha|$ between BSBM model implementation with and without contribution of $\psi$ to the Friedmann equation, as a function of redshift $z$. Here $\zeta_{m}=10^{-8}$ and $\omega=1$.}
\label{fig:friedif_redshift}
\end{figure*}

\label{sec:bsbm_subtle}
The equations of motion used earlier (and throughout the BSBM literature) do not conserve energy. In particular, using the fact that the field energy density is $\overline{\omega}\dot{\psi}^{2}/2$ and the field equation of motion [Eq.~(\ref{eq:time_dependent})], we see that $\dot{\rho}_{\psi}=-3\overline{\omega}\dot{\psi}^2H-2e^{-2\psi}\zeta_{m}\dot{\psi}$. Given that a field with no potential has equation-of-state parameter $w_{\psi}=1$, this can be rewritten as

\begin{equation}
\dot{\rho}_{\psi}+3(1+w_{\psi})\rho_{\psi}=-2\sqrt{\frac{2}{\overline{\omega}}}e^{-2\psi}\zeta_{m}\rho_{m}\sqrt{\rho}_{\psi}.\label{eq:bsbm_edens}
\end{equation}

The left-hand side of Eq.~(\ref{eq:bsbm_edens}) is of course the standard term for energy flow out a fixed physical volume in an expanding universe. The right-hand side is an energy flow term from the scalar field into matter.

The usual $\Lambda$CDM continuity equation for matter density,
\begin{equation}
    \dot{\rho}_m + 3H \rho_m = 0,
\end{equation} must acquire an additional term if energy conservation is to hold, and so we have that

\begin{equation}
    \dot{\rho}_m + 3H \rho_m - 2\sqrt{\frac{2}{\overline{\omega}}} e^{-2\psi} \zeta_{m} \rho_m \sqrt{\rho_\psi} = 0.\label{eq:rhom_mod}
\end{equation}

There is nothing \textit{ad hoc} about using Eq.~(\ref{eq:rhom_mod}). The additional term on the right-hand side of Eq.~(\ref{eq:bsbm_edens}) comes from the substitution $\mathcal{L}_{\rm em}\to \zeta \rho_{\rm m}$ - this amounts to saying that the electrostatic influence of charged particles contributing to $\mathcal{L}_{\rm em}$ leads to an additional interaction between $\psi$ and matter, in some sense, integrating out the relevant electromagnetic fields. These fields should backreact on the plasma that sources them, and if we are to take this Lagrangian substitution at face value, the Bianchi identity implies the $\nabla_{\mu}\left(e^{-2\psi}T^{\mu \nu,{\rm matter}}\right)=0$. This can be readily applied in a FRW (Friedmann-Robertson-Walker)cosmology to obtain Eq.~(\ref{eq:rhom_mod}).

This modified equation for matter density and the second-order differential equation for the scalar field can then be written as a system of three first-order ordinary differential equations as follows:

\begin{equation}
    \frac{df}{dz}=\frac{3}{1+z}f+\frac{2}{g(1+z)}e^{-2\psi}\zeta_{m} f u,
\end{equation}
\begin{equation}
    \frac{d\psi}{dz}=-\frac{u}{g(1+z)},
\end{equation}
\begin{equation}
    \frac{du}{dz}=\frac{3}{1+z}u+\frac{6\Omega_{m,0}}{g(1+z)}\frac{1}{{\omega}}e^{-2\psi}\zeta_m f
\end{equation}

where $f(z)=\Omega_m(z)/\Omega_{m,0}$, $u(z)=-\dot{\psi}/[H_0 g(1+z)]$ is a dimensionless velocity for the scalar field and
\begin{widetext}
\begin{equation}
g^2=\Omega_{m,0}f\left(1+|\zeta_{m}|e^{-2\psi}\right)+\Omega_{r,0}(1+z)^4e^{-2\psi}+\frac{{\omega} }{6}u^2+\Omega_{\Lambda,0}\label{eq:hubap}
\end{equation}
\end{widetext}
represents the modified dimensionless Hubble parameter.

This system of differential equations was then solved numerically over a range of redshifts and compared to the non energy-conserving implementation with $\zeta_m=10^{-8}$ and ${\omega}=1$. The left panel of Fig.~\ref{fig:econserve_redshift} shows the fractional difference in the evolution of $\Delta\alpha/\alpha$ over $z$ for these two models. Agreement is sufficient for our purposes (a $\sim 10\%$ correction to our constraints). We also compared $\Delta\alpha/\alpha$ values at recombination for a range of $\zeta_m$ values from $-10^{-8}$ to $10^{-8}$ (roughly the constraint from \emph{Planck}),  with $\omega$ still set to $1$. 
 The results are displayed in the right panel of Fig.~ \ref{fig:econserve_redshift}, which shows that the two calculations agree well over this range of $\zeta_m$, and vanish when $\zeta_{m}\to 0$. The agreement improves with even higher $|\zeta_{m}|$.

It is worth pausing to consider the interpretation of these corrections. In particular, we can state the matter continuity equation as
\begin{equation}
    \dot{\rho}_m + 3H \rho_m =  -2 e^{-2\psi} \zeta_{m}\dot{\psi} \rho_{m}.\label{eq:matter_cont}
\end{equation}

With the ${ansatz}$ $\rho\equiv a^{-3}f(a)$, we find (applying separation of variables) that 
\begin{equation}
    \rho_{m}=\frac{C}{a^{3}}e^{\zeta_{m}e^{-2\psi}}.
\end{equation}
This scaling suggests the definition $\tilde{\rho}_{m}\equiv \rho_{m}e^{-\zeta_m e^{-2\psi}}$ as a physical matter density with the right redshift dependence. The matter-dependent term in the Friedmann equation (in the BSBM model) is \cite{Sandvik:2001rv,Barrow:2002zh}
\begin{align}
    H^2_{m}=&\frac{8\pi G}{3} \left\{\rho_m+\zeta_m e^{2\psi}\rho_m \right\}\\=&
    \frac{8\pi G}{3}e^{\zeta_m e^{-2\psi}}\left[1+\zeta_m e^{2\psi}\right] \tilde{\rho}_{m}\\=&\frac{8\pi G\tilde{\rho}_{m}}{3}\frac{m_{m}(\psi)}{m_{m}(\psi=0)}.
\end{align}
where $\tilde{m}_{m}$ can be interpreted as the $\psi$-modulated  mass of nonrelativistic particles (e.g. baryons, dark matter). If $\rho_{m}$ is treated (erroneously) as a substance that redshifts as $a^{-3}$, then this amounts to the approximation $m_{m}(\psi)=m_{m}(\psi=0)$. The analysis above shows that the error in $\alpha$ evolution induced by this approximation in the BSBM is negligible.

This effective modulation of the mass of nonrelativistic particles is a well-known feature of theories with nonminimally coupled scalar fields, in which the relevant matter fields travel on geodesics of a different metric than the one satisfying Einstein's equations (see e.g. Refs.~\cite{Karwal:2021vpk,Lin:2022phm} for recent applications to early dark energy phenomenology and Refs.~\cite{Bean:2000zm,Bean:2007ny,Bean:2008ac,Miranda:2017rdk} for earlier applications).

In Sec.~\ref{sec:sf_dynamics}, the contribution of $\psi$ to the Hubble expansion itself was not included. The dynamics modeled above do include this contribution, and we have verified that the overall correction to $\Delta \alpha/\alpha$ does not affect the remaining results of this paper. It is instructive to consider the contribution of $\psi$ to the Hubble expansion (separate and apart from the energy nonconservation or modulated scalar mass effect discussed above). The results are shown in Fig.~\ref{fig:friedif_redshift}. We see that this effect (taken alone) is even smaller than the energy conservation correction.

Additionally, Eq.~(\ref{eq:hubap}) is only valid for massless neutrinos, whereas the fiducial cosmology used in \emph{Planck} data analysis assumes a single neutrino with $m_{\nu}=0.06~{\rm eV}.$  Using the equations in Ref.~\cite{WMAP:2010qai}, which correct the Hubble factor analytically for the contribution of massive neutrinos (through their full relativistic to nonrelativistic transition), we assess if neutrinos induce any change to the values of $\alpha$ determined in the preceding portions of this section. We find a fractional error in $\Delta \alpha/\alpha$ at $z=1100$ of $\sim 10^{-3}$, meaning that this effect does not alter any of the conclusions of this paper.

\end{appendix}

\bibliography{main3}

\end{document}